\documentclass[sigconf, nonacm]{acmart}

\newcommand\vldbdoi{XX.XX/XXX.XX}

\newcommand\vldbavailabilityurl{https://github.com/saranyacsekar/ElasticScheduler}
\newcommand\vldbpagestyle{plain} 
\newcommand{\fullversion}[1]{}

\usepackage{graphicx}
\usepackage{textcomp}
\usepackage{xcolor}

\usepackage{comment}
\usepackage{mathtools}
\usepackage{amsmath}

\usepackage{calc}
\usepackage{color,soul}
\usepackage{balance}

\usepackage{algorithm}
\usepackage[noend]{algorithmic}
\usepackage{multicol}
\usepackage[toc,page]{appendix}
\usepackage{balance}
\usepackage{subcaption}
\usepackage{multirow}

\usepackage[latin1]{inputenc}

\begin{document}
\title{Elastic Scheduling of Intermittent Query Processing in a \\ Cluster Environment}


\author{Saranya Chandrasekaran}
\orcid{0009-0005-3960-7824}
\affiliation{%
  \institution{U R Rao Satellite Centre, ISRO}
  \institution{IIT Bombay}
  \postcode{400076}
}
\email{saranyac@cse.iitb.ac.in}

\author{S. Sudarshan}
\orcid{0000-0002-6230-2288}
\affiliation{%
  \institution{IIT Bombay}
  \streetaddress{}
  \city{}
  \state{}
  \postcode{400076}
}
\email{sudarsha@cse.iitb.ac.in}


\begin{abstract}
Many applications process a stream of tuples over a window duration and require the results within a specified deadline after the end of the window.  
For such scenarios, processing tuples intermittently (in batches) instead of eagerly processing tuples as they arrive significantly reduces the overall cost. Earlier work on intermittent query processing has addressed only fixed environments. 

In this paper, we propose scheduling schemes for batched processing of tuples in an elastic parallel environment, scaling nodes up or down.
Our scheduling schemes ensure meeting the deadlines while incurring minimum cost. Our schemes also handle multiple concurrent queries, the arrival of new queries, and input rate variations. 

We have implemented our schemes on top of Apache Spark, in the AWS EMR environment, and evaluated performance with both TPC-H and Yahoo Streaming datasets.  Our experimental results show that our scheduling algorithms significantly outperform alternatives, such as using a fixed set of nodes without elasticity, or using
Spark streaming. 
\end{abstract}

\maketitle

\pagestyle{\vldbpagestyle}



\section{Introduction}
Many applications have a set of analysis queries that are run on a regular basis, where results are needed only at deadlines and not required on per tuple basis. 

Stream processing engines such as Apache Spark and Apache Flink update the aggregate/status as and when a new tuple arrives or when a micro batch of tuples arrives. This method of processing eagerly can lead to significant overhead.

Instead, the fact that results are needed only by a deadline can be utilized to process tuples in batches, thereby reducing overhead. If there is sufficient time to process all tuples together, starting after the window end, then a single batch is sufficient to complete the processing. However, in general, there may not be sufficient time to process the entire data in the gap between the end of the window and the query deadline.

In such cases, tuples may be collected and processed in batches, and then the partial aggregates on each batch can be further aggregated to get the final result. 
Tang et al.\ \cite{ref_iqp} refer to this method of processing queries in parts and combining the intermediate results as \textit{Intermittent Query Processing}. 



Our earlier work \cite{ref_self_ideas} presented scheduling strategies for intermittent query processing that minimize cost for an environment with fixed nodes, while ensuring deadlines are met.  Experimental results with the TPC-H Dataset and queries presented in  \cite{ref_self_ideas} show that the scheduling schemes achieve significant cost reduction compared to processing using Spark Streaming with microbatches. 

To handle large ingestion rates, we need to use parallel processing. Processing in batches means that in between batch executions, the system may be idle, waiting for enough tuples to arrive.
Further, in case there is a spike in the load or the input rate increases, with a fixed set of nodes, deadlines may get missed.


A parallel elastic platform where nodes can be added or removed based on demand
can be used to address both the above-mentioned problems, reducing cost, while being able to scale up to handle load spikes. While fixed resource scheduling, as in \cite{ref_self_ideas}, requires a large number of nodes to handle stringent deadlines or higher input rates, our present work proposes an elastic scheduling methodology which initially uses a default number of nodes and further adds or removes nodes based on demand. 
The Elastic Map Reduce (EMR) platform provided by Amazon Web Services (AWS) is an example of such a platform where billing is done on per second basis based on the resource usage.

Our performance results in Section \ref{sec_results_fixedConfig} show that using an elastic configuration can significantly reduce cost compared to a minimal fixed configuration that can meet the deadline.

In this paper, we address the scheduling of Intermittent query processing in a parallel elastic platform. While fewer nodes may result in a longer processing time, the overall cost incurred may be cheaper compared to processing with a larger number of nodes, but at the risk of missing deadlines. 
At the other end, processing all tuples together in a single batch after the window end may require a large number of nodes to meet the deadline, thereby incurring a very high cost.  

Thus, given a set of tuples to be processed, the overall cost required for processing the same varies depending on the number of nodes used, and their associated node configuration cost.  While the
reduction in processing duration as the number of nodes increases is typically linear or sub-linear, the cost incurred typically increases linearly or super-linearly as the number of nodes increases.  

Our goal is to schedule the execution of tuple batches at appropriate times on an appropriate number of nodes to minimize the cost while meeting deadlines.
We note that a batch can be scheduled only after all the tuples in the batch have arrived.


Further, requirements may vary dynamically due to varying input rates, as well as the addition of new queries to the system, requiring dynamic changes to the schedule.



In this paper, we propose an elastic scheduling approach that is based on a simulation model. Given a set of queries, the input rate and initial configuration, the simulation model is used to estimate the appropriate batch size and number of nodes required over different time points to meet the deadline.  Our approach is to run the simulation model for different configurations and choose the one that incurs the least cost while ensuring the deadline constraint is met. 

Based on the simulation results, our scheduler does the node management by issuing appropriate node resize requests. In general, the addition or removal of nodes in an elastic platform takes a few minutes to set up or release. Our scheduler handles this delay to ensure that the required number of nodes are available when needed.

Our contributions are as follows
\begin{enumerate}
\item We propose strategies to derive the cost model for different configurations.

\item We have devised a simulation model (in Section~\ref{sec_schgeneration} and \ref{sec_schoptimization}) to generate the node requirements based on the tuple arrival rate and cost model of the query.  
\item We present the query schedule methodology to determine the best configuration along with the appropriate batch size, in Section \ref{sec_configdetermination}.
\item We describe how to schedule execution with the addition and deletion of nodes in Section \ref{sec_schexecution}.  We describe how to handle variable input rates (in Section \ref{sec_variableIpRate}) and how to handle partial aggregation in scheduling (Section \ref{sec_partial}). 
\item The scheduling strategies have been implemented on Apache Spark, executed on AWS EMR; details are presented in Section \ref{sec_implementation}.
\item We present experiments with two benchmark datasets: TPC-H and Yahoo Streaming dataset, including different input rate profiles and deadline scenarios.  The results show the significant benefits of our approach compared to alternatives (Section \ref{sec_results}).
\end{enumerate}

Related work is discussed in Section~\ref{relwork}, while
Section~\ref{sec:concl} summarizes the conclusions and future work.


\section{Background}
\label{sec_need_for_scheduling}

In this section, we give an overview of the problem description and then explain the factors that affect query scheduling.

\subsection{Problem Description} \label{sec_probdesc}

We consider the problem of processing queries with deadlines, on streaming data, intermittently, on an elastic cloud platform. 
A batch of input tuples for a query 
is processed at a time, in parallel across multiple nodes. 
We assume only one query is processed at a time; 
extensions to consider schedules with  concurrent processing of multiple queries are part of future work.

An elastic schedule is a sequence of entries, where each entry consists of the following 4 attributes: (query ID,  batch start time, batch size, number of nodes), which represents the execution of the selected query on a batch of specified size, on the specified number of nodes. 
Our goal is to derive an elastic schedule that incurs the minimum cost, while meeting query deadlines.

The number of alternative configurations (number of nodes) is quite large.  To minimize the optimization cost, we consider only some fixed set of configurations, for example, configurations with 2, 4, 10, 14, and 20 nodes.
Let these configurations be represented as $Cf_i$ where i varies from 1 to n, s.t. each configuration $Cf_{i+1}$ has more nodes than $Cf_i$.

Typically, multiple queries need to be processed.
We assume that the queries are independent of each other, each with its own deadline.   

We consider queries that are incremental in nature, allowing tuples to be processed in batches, i.e., incrementally, and combining
intermediate results, subsequently, to get the final results; this property is  referred to as incrementability in \cite{ref_thriftyqp}.    

Queries often run on a single input stream, 
but may access multiple stored or static relations. 
We assume that the static data does not change over the period of query execution. Hence, each input stream batch is joined against the static data to get the join results. 

Queries that run on multiple streams require stream-to-stream joins.  In general, tuples in one batch of a stream could potentially match tuples from any other batch of the stream, which can be very expensive.
Similar to \cite{ref_self_ideas}, for simplicity, our implementation assumes that batches of all streams are aligned, and tuples in a batch of one stream can
only match tuples in the same batch of any other stream. For example, consider the join between orders and lineitem from the TPC-H Dataset. It is reasonable to expect the order and its associated lineitem tuples to arrive in the same batch. However, details on handling late arriving tuples are covered in Section \ref{sec_variableIpRate}.


Since many analysis queries are recurrent in nature, the cost model of a query can be learnt from the past execution logs, as explained in Section \ref{sec_cost_model_exp}. The input rate can be modeled as uniform or non-uniform. While the uniform input rate model assumes that the number of tuples arriving per second does not vary, the non-uniform input rate model can accommodate changing arrival rates, for example, peak versus non-peak hour traffic.  Initially, we consider the input rate is known ahead of time, and later in Section~\ref{sec_variableIpRate} we address handling dynamically varying input rate.

\fullversion{Given a single node configuration, while processing all tuples in a single batch incurs lesser cost than processing in multiple batches, there may be cases where deadlines are stringent or the requirements may be dynamic in nature, where processing cannot be postponed until all tuples arrive.}

\fullversion{We measured experimentally the time taken for processing 25 GB of TPC-H dataset for query TPC-Q1 in an EMR cluster with the different configurations such as 1P-1C-1T, 1P-1C-3T, 1P-1C-9T, 1P-1C-13T and 1P-1C-19T where P denotes the Primary node, C denotes Core node and T denotes Task node. While there is a single Primary node in all the configurations, the number of Core and Task nodes varies as mentioned above. We observed that the processing duration approximately reduces by half on doubling the number of nodes for 1P-1C-1T, 1P-1C-3T and 1P-1C-9T. But further adding nodes i.e. for 1P-1C-13T and 1P-1C-19T configurations there is no significant reduction in processing time.}


\subsection{Factors affecting Query Scheduling}
\label{ssec:sched:factors}

The parameters of a query that affect scheduling decisions and relevant notations are given in Table \ref{tbl:query_attributes}. The main decisions to be made as part of query scheduling are the batch sizes, their scheduling time points, and the number of nodes required over the time period, with the goal of minimizing the overall computation cost, while satisfying the deadline constraints.

\begin{table}
\caption{Notation for Query Parameters}
\begin{center}
\raggedright
\begin{tabular}{lp{2.2in}}
\toprule
\textbf{Notation}&\textbf{Description} \\
\midrule
    \textit{queryID} & Unique Identifier for the Query \\ 
    \textit{windStartTime} & Time at which tuple arrival starts  \\  
    \textit{windEndTime} & Time at which tuple arrival stops \\  
    \textit{deadline} & Time by which the Query processing must be completed   \\ 
    \textit{inputStream} & Denotes the input stream of the query \\     
    \textit{inputRate} & {Rate at which tuples arrive for inputStream (for fixed input rate case)}\\ 
    \textit{numTupleTotal} & {Denotes the total number of tuples to be processed}  \\ 
    \textit{minCompDur} & {Time period for processing all the tuples in a single batch} \\ 
    \textit{minCompCost} & {Monetary Cost for processing all the tuples in a single batch} \\ 
    \textit{slackTime} & {The maximum time  beyond  which the query processing cannot be further delayed without missing the deadline}\\
\bottomrule
\end{tabular}
\label{tbl:query_attributes}
\end{center}
\end{table}

In order to determine the best cluster configuration that incurs the least cost while meeting query deadlines, we have to model the duration and the monetary cost for processing a given number of tuples. The cost model for time and monetary cost may be linear or non linear.  In our implementation, we use a simple linear cost model combining two factors: the processing cost per tuple,  and the per-batch overhead cost. 

The time taken for processing a given number of tuples depends on the number of nodes used and the number of batches into which the input is divided. The amount of time taken for processing data in a parallel manner is given by Amdahl's law as in equation \ref{eqn_tparallel}

\begin{equation} \label{eqn_tparallel}
   T_{P}=((1-P)+P/N_{p})*T_{S}+O_{N}    
\end{equation}

where $T_{S}$ denotes the time to run an application in serial version, P represents the fraction of work that can be done in parallel, $N_{p}$ denotes the number of processors, and $O_{N}$ denotes parallel overhead in using N threads.  

$T_{S}$ is dependent on $N_{t}$, the number of tuples, and CPT, the Cost Per Tuple. Also, there is an overhead cost, $O_{X}$, when the processing is done in batches. The cost model in terms of cost per tuple and overhead cost is given in equation \ref{eqn_cost_model_parallel}

\begin{equation} \label{eqn_cost_model_parallel}
     T_{P}=((1-P)+P/N_{p})*N_{t}*CPT+O_{N}+O_{X}   
\end{equation}

In general, we process tuples in batches, numbered $1 \ldots B$,
each with its own number of tuples $t_{i}$, and number of nodes $p_{i}$. Let
$BatchTime_q(p_i, t_i)$ denote
the execution time for query $q$ on $p_i$ nodes on a batch of $t_i$ tuples.  Similarly, let the cost of executing  query $q$ on $p_i$ nodes on a batch of $t_i$ tuples, be 
denoted as $BatchCost_q(p_{i}, t_{i})$. 
Once all batches are processed, the final aggregation is done using $p$ nodes. The total cost of processing the batches, including the final aggregation cost, $FinalAggCost_q$, is then defined as in Equation \ref{eqn_cost_model_func} below. 



\begin{equation} \label{eqn_cost_model_func}
    TotalCost_q=\sum_{i=1}^{B}BatchCost_q(p_{i},t_{i}) + FinalAggCost_q(p)
\end{equation}

 
We execute each query with different  batch sizes on different number of nodes and learn the functions $BatchTime_q()$, $BatchCost_q()$, $FinalAggTime_q(p)$ and $FinalAggCost_q(p)$ by fitting these measurements to a linear or piecewise linear cost model (see Section \ref{sec:elasticsch_costmodel} for  details). 



Using the cost models, we generate an elastic schedule which contains the number of nodes required for processing the query batches at different points of time such that the cost incurred in minimised while meeting the deadlines. 

In an elastic setting, provisioning (adding) a node may take some time (up to several minutes on AWS EMR).
This provisioning time needs to be taken into account when deciding on a schedule, and the request should be sent ahead of when it is needed; more details are provided in Section \ref{sec_schexecution}.

\fullversion{
\section{Single Query Static Scenario} \label{sec_static}

In the static scenario we consider a single query whose input rate is as per the model and the scheduler is aware of the query requirement. Scheduling for a single query under the static scenario is presented in this section. 

Consider a query with an input rate of 1 tuple per sec, whose window start time and window end time be 1 and 12 respectively. Let the deadline be 16 and the cluster configurations be $C_{1}$ and $C_{2}$ whose cost-per-second is 1 and 2.5 dollars respectively. Here, for the sake of simplicity, a linear cost model without overhead cost is assumed,  where 2 tuples can be processed per second in $C_{1}$ and 4 tuples can be processed per second in $C_{2}$. Let the final aggregation cost be negligible.   

Here, processing in a single batch is possible only with $C_{2}.$ All tuples can be batched together and processed in 3 seconds, starting after the window end not later than 13 seconds. The cost incurred for this will be 7.5 dollars. On the other hand processing with $C_{1}$ is feasible provided tuples are processed in multiple batches. The last batch can be processed only after window end, i.e. starting at time point 12 sec. As 2 tuples can be processed per second, 8 tuples can be processed in the last batch. Rest of the tuples must be processed prior to the window end. The pending tuples are available at time point 4 and hence these can be processed in a batch starting at any time after 4 seconds and completed within the window end. Thus the tuples can be processed in 2 batches with a total duration of 6 seconds and incurring a cost of 6 dollars. 

Processing in single batch, especially for stringent deadlines may require more nodes. On the other hand with fewer nodes processing can be done in smaller batches, but processing has to be started earlier. Thus the goal of scheduling is to find the ideal number of nodes which completes the query processing with the least monetary cost while meeting its deadline.


\floatname{algorithm}{Function}
\begin{algorithm}
\begin{flushleft}
\caption{SingleQryCostEst($C_i$) \label{algo_single_qry_estBatchSize_cluster}} 
\begin{algorithmic}[1]
\FOR{$(numBatch=1;numBatch<MAXBATCHES;numBatch++)$}
\STATE $modDeadline$ $=$ $q.deadline$
\IF{$(numBatch!=1)$}
\STATE $aggCost=EstAggCost(numBatch,C_i)$
\STATE $modDeadline$ $-=$ $aggCost$
\ENDIF
\STATE $batchLength$ $=$ $0$
\STATE $totalCost$ $=$ $0$
\STATE $q.numTuplePending$ $=$ $q.numTupleTotal$
\STATE $timePt$ $=$ $modDeadline$
\WHILE{$(q.numTuplePending > 0)$}	 
\STATE $ipAvailTime=InputTime(q.inputStream,$ $q.numTuplePending)$
\STATE $timeDur=timePt-ipAvailTime$
\STATE $numTupleProc,cost=EstimateTuplesProcessed(timeDur,C_i)$
\IF{$(numTupleProc>q.numTuplePending)$}
\STATE $numTupleProc=q.numTuplePending$
\STATE $timeDur,cost=EstimateDuration(q.numTuplePending,C_i)$
\ENDIF
\STATE $Batches.StartTime[batchLength] = timePt-timeDur$
\STATE $Batches.tuplesProcessed[batchLength]=numTupleProc$
\STATE $totalCost+=$ $cost$
\STATE $q.numTuplePending$ $-=$ $numTupleProc$
\STATE $timePt$ $-=$ $timeDur$
\ENDWHILE
\IF{($batchLength$ $<=$ $numBatch$)}
\STATE $RETURN$ $totalCost,Batches$
\ENDIF
\ENDFOR
\end{algorithmic}
\end{flushleft}
\end{algorithm}

Function \ref{algo_single_qry_estBatchSize_cluster}, \textit{SingleQryCostEst},  gives the methodology to compute the number of batches, the batch sizes and the cost for processing the query for a given configuration. The scheduling algorithms are implemented in a custom scheduler and the implementation details are explained in Section \ref{sec_implementation}. Initially we assume that all the tuples can be processed in a single batch. The last batch can be scheduled not earlier than window end and the number of tuples that can be processed in the last batch is determined based on the duration between the window end and deadline. The pending tuples have to be processed prior to window end and the batch processing cannot be started until the input available time is met. Similarly the pending tuples have to be processed in one or more batches. The function \textit{InputTime} gives the time at which the pending tuples will be available as per the input profile for the given stream. The function \textit{EstimateTuplesProcessed} determines the number of tuples that can be processed for the given duration and configuration,  and its monetary cost. The function \textit{EstimateDuration} gives the duration and the cost for processing the given number of tuples with the said configuration. If the total number of batches required for processing is greater than the initial assumption, then we increment the number of batches and re-compute the batches and cost by accommodating the aggregation cost in the deadline, until the number of batches required for processing is less than or equal to the initial assumption. For each configuration we determine the number of batches, batch sizes, their duration and its cost. Then we select the optimal configuration which incurs the least cost while meeting the deadline. 


The optimal configuration and its schedule can be computed in offline mode as input rate and query attributes are known apriori. Then, based on the schedule generated, the required cluster set up can be initialised and query processing can be scheduled using the generated schedule. Also as a cluster set-up requires some considerable time, computing the schedule offline and bringing up the cluster prior to the actual schedule will ensure that the cluster initialisation delay does not affect deadlines.
 }

We explain our scheduling strategy in the subsequent sections. We consider a non-preemptive scheduler that performs a set of tasks periodically. These tasks include running a simulation model to generate a schedule and executing the generated schedule. We explain schedule generation in Section 3, followed by schedule execution in Section 4. Section 5 discusses handling of variable input rates.


\section{Execution Simulation} \label{sec_simulation}

One of the main functionalities of the scheduler is to assess the sufficiency of the currently allocated nodes for meeting the query deadline, and to generate a schedule to acquire or release nodes. To generate the schedule, the scheduler simulates the execution or processing of the given set of queries based on batch readiness times, for the given batch sizes and number of nodes. 

All algorithms in this section perform simulation; actual scheduling of query batches for execution is discussed later in Section~\ref{sec_schexecution}.

We first explain the generation of a schedule for a given batch size and initial configuration in Section 3.1, and the optimization of the generated schedule in Section 3.2. Section 3.3 describes the simulation process for different batch sizes and initial configuration values to determine the least-cost schedule.


\subsection{Schedule Generation} \label{sec_schgeneration}

We consider a non-preemptive scheduler where once a batch execution begins, it is allowed to complete without interruption.
We upper bound the batch size for each query, such that each batch can be processed within a maximum duration called $C_{max}$. 
This ensures that any new query 
can be considered for processing within an interval $C_{max}$ (plus the simulation time) from its arrival.


We consider configurations with at least 2 nodes.
The minimum time for processing the input of $N$ tuples would be by using a single batch of size $N$.
With 2 nodes, this time is  $BatchTime(2,N)$.
To minimize overheads due to small batches, we choose a batch size, $X$, which is the minimum  value of x such that $\lceil N/x \rceil * BatchTime(2,x)$ $\leq$ 2 $*$ $BatchTime(2,N)$. 
We upper bound the batch size to the maximum batch size $x$ such that $BatchTime(2,x) < C_{max}$. 


The batch size determined based on the above scheme is referred to as batch size factor 1X. Further, for simulation, we vary the batch size factor in multiples of 1X, i.e. for a batch size factor of 2X, the batch size is computed as 2 times 1X. The simulation is repeated for different configurations, i.e. number of nodes, say $Cf_{i}$ to $Cf_{n}$ and batch size factors, say from 1X to SX, as explained in Section \ref{sec_configdetermination}. 

\textit{GenSchedule} generates the schedule based on execution simulation for the given configuration and batch size factor, provided it is feasible, by invoking \textit{GenBatchSchedule}, as explained next. 

\subsubsection{\textbf{Algorithm GenSchedule:}}

Algorithm~\ref{algo_multi_qry_simulate} shows the GenSchedule() function.
The function takes as input
\textit{QList}, which is the input query list.
Each query has associated query attributes as listed in Table \ref{tbl:query_attributes}. In addition, \textit{QList} maintains with each query the number of tuples pending to be processed for the query. \textit{simuQList} represents the query list used for simulation, and it is initialized to \textit{QList} in line 2 of Algorithm \ref{algo_multi_qry_simulate}. 

The variable \textit{qryBatchSch} contains the list of query batches that are scheduled, with associated information for each batch, such as the time, query ID, batch size, batch computation start and end times, pending tuples and most importantly, the number of nodes required for processing that batch. Initially, \textit{qryBatchSch} contains a single entry with the simulation start time and the initial number of nodes as in lines 6 and 7 of Algorithm \ref{algo_multi_qry_simulate}.

The \textit{GenSchedule}() function invokes \textit{GenBatchSchedule()} to generate the query batch schedule for a specific batch size factor and number of nodes. It returns true if all query batches could be scheduled successfully, else it returns false. 

The variable \textit{schStartIndex} tracks the current point from which the simulation has to be carried out in \textit{GenBatchSchedule}. Initially, it is set to 0.  If \textit{GenBatchSchedule} returns true, then cost is computed based on the number of nodes and their usage and \textit{GenSchedule} function returns the cost and the schedule, \textit{qryBatchSch}.

If \textit{GenBatchSchedule} returns false, i.e. the slack time of a query is negative, then we add additional nodes to earlier query batches. The intention to add additional nodes to earlier query batches is to complete these batches earlier, thereby getting sufficient slack for the current query batch (the one with negative slack time with the current schedule) with fewer nodes than requiring a higher number of nodes for speeding up only the current batch.

Then the simulation is repeated, where \textit{GenSchedule} invokes the function \textit{GenBatchSchedule},  where the simulation is rerun from the last query batch by setting it to the next configuration, i.e. $Cf_{i+1}$. If the slack time still remains negative, then the simulation is run with a higher configuration for the successively earlier batches until a positive slack time is achieved, or we reach the first batch. On reaching the first batch, if the slack time is still negative, we rerun the process above with the next higher configuration, starting from the last batch backwards. This process continues until a feasible schedule is obtained, or if a feasible schedule is not possible even on reaching the maximum number of nodes, then the simulation returns an empty schedule.

Lines 15 to 25 in Algorithm \ref{algo_multi_qry_simulate} show the steps followed when \textit{GenBatchSchedule} returns false.
When \textit{schStartIndex} becomes negative, or if there exists idle time between the current batch end time and the subsequent batch start time, then the \textit{schStartIndex} is set to the last entry in \textit{qryBatchSch}, and the number of nodes is incremented, thereby retrying the simulation with a higher number of nodes.


While invoking \textit{GenBatchSchedule}, the simulation is carried out from a specific time point which corresponds to the \textit{BET}, i.e. batch end time, of the last scheduled query batch w.r.t. \textit{schStartIndex} in \textit{qryBatchSch}.  
The number of pending tuples for each query at any given time point is obtained from the last completed query batches of that query, scheduled in \textit{qryBatchSch} w.r.t \textit{schStartIndex}. While invoking \textit{GenBatchSchedule}, \textit{simuTime} is set to \\ \textit{qryBatchSch[schStartIndex-1].BET}, along with the query attributes corresponding to the last query batches for each of the queries (lines 21 to 25 in Algorithm \ref{algo_multi_qry_simulate}). Then, the simulation is rerun only for the subsequent query batches.  

Simulation time can be reduced by decrementing \textit{schStartIndex} in 
steps of some $K$, instead of decrementing in steps of 1 in Algorithm \ref{algo_multi_qry_simulate}, 
thereby trying fewer alternative node schedules. More details on this are discussed in Section \ref{sec:elasticsch_optimality}.


\textit{GenBatchSchedule} is invoked until the \textit{exitFlag} becomes true, which occurs when a successful schedule gets generated, or if \textit{numNodes} has exceeded the maximum number of nodes, in which case \textit{GenSchedule} returns with an empty schedule.


\floatname{algorithm}{Algorithm}
\begin{algorithm}
\caption{\textit{GenSchedule (initNumNodes, batchSizeFactor, QList[], simuStartTime)}} \label{algo_multi_qry_simulate}
\begin{flushleft}
\begin{algorithmic}[1]
\STATE \textbf{Output:} \textit{cost, qryBatchSch [time, qryID, batchSize, BST, BET, reqNodes, pendingTuples]} \\
\textit{//qryBatchSch is a list of query batch schedules with time, query ID, batch size, batch computation start and end time, pending tuples, required number of nodes}
\STATE simuQList = QList
\STATE numNodes = initNumNodes
\STATE simuTime = simuStartTime
\STATE schStartIndex = 0
\STATE qryBatchSch[schStartIndex].time = simuTime
\STATE qryBatchSch[schStartIndex].reqNodes = numNodes
\STATE exitFlag = false
\WHILE{(exitFlag == false)}
\STATE posSlack = GenBatchSchedule (simuQList, \&qryBatchSch, batchSizeFactor, $C_{max}$, simuTime, schStartIndex, numNodes)
\IF{(posSlack)}
\STATE \textit{//Schedule generated successfully}
\\ Compute cost from qryBatchSch based on number of nodes required and its duration of usage
\STATE exitFlag = true
\ELSE
\STATE schStartIndex = schStartIndex-1 \textit{//can be varied in steps of K instead of 1 to optimize the time taken for simulation}
\IF{(schStartIndex < 0 OR (qryBatchSch[schStartIndex+1].BST - qryBatchSch[schStartIndex].BET > 0))}
\STATE schStartIndex = qryBatchSch.length-1
\STATE numNodes++
\IF{(numNodes > MAXNODES)}
\STATE Return Empty Schedule \textit{//Schedule not possible}
\ENDIF
\ENDIF
\IF{schStartIndex == 0}
\STATE simuTime = simuStartTime
\ELSE
\STATE simuTime = qryBatchSch[schStartIndex-1].BET
\ENDIF
\STATE Update pendingTuples in simuQList from qryBatchSch
\ENDIF
\ENDWHILE
\STATE \textbf{end while}
\STATE \textbf{Return} cost, qryBatchSch
\end{algorithmic}
\end{flushleft}
\end{algorithm}

\subsubsection{\textbf{Algorithm GenBatchSchedule:}} Function
\textit{GenBatchSchedule()} simulates the execution of query batches for the set of input queries from \textit{GenSchedule} with the given batch size factor and number of nodes. The simulation is carried out from \textit{schStartIndex} in \textit{qryBatchSch}. 

\textit{SimuTime} is initialized to the input value of \textit{simuStartTime}, when the simulation starts in line 2 of Algorithm \ref{algo_generate_schedule}. 

The scheduler determines the actual batch of each query based on the batch size factor and $C_{max}$. Then, the scheduler determines the next batch readiness time (i.e. the batch to be scheduled), $NextBRT$, for each query based on the batch size and the estimated input rate. Let $BCT$ denote its Batch Computation Time and $FAT$ denote the Final Aggregation Time. $BCT$ and $FAT$ are computed using the cost model of the corresponding configuration, $Cf_{i}$, where $i$ is based on the number of nodes passed as input to the Function GenBatchSchedule()



\begin{equation} \label{eqn_bst}
Q_{i}.BST=
\begin{cases}
   \textit{SimuTime} & \text{if }  \textit{SimuTime} > Q_{i}.NextBRT \\
    Q_{i}.NextBRT  & \text{else }      
\end{cases}
\end{equation}

Let $BST$ and $BET$ denote the Batch Start Time and Batch End Time, respectively, corresponding to each query batch execution. 
Let $Q_i$ denote the $i^{th}$ batch of query $Q$.
BST for $Q_i$ is determined as in Equation \ref{eqn_bst}, and then the slack time of the query at the start of the $i$th batch is computed, as in Equation \ref{eqn_slack_single_qry}; here $n$ denotes the total number of batches.
    
 

\begin{equation} \label{eqn_slack_single_qry}
        Q.SQSlacktime = Q.Deadline - Q_{i}.BST - ({\sum_{j=i}^{n}Q_{j}.BCT + Q.FAT)}
\end{equation}

If query batches exist that are ready at \textit{simuTime}, i.e. whose \textit{NextBRT} is equal to or prior to \textit{simuTime}, then the query with the least slack time among such query batches is scheduled for processing (in the simulation). If none of the query batches are ready at \textit{simuTime}, then sorting is based on the minimum lexicographic order of \textit{(NextBRT, SQSlacktime)} (line 11 in Algorithm \ref{algo_generate_schedule}). The query batch thus selected is scheduled for processing, provided it has positive slack time. 

This ensures that we choose the query with the least slack time, thus using Least Laxity First (LLF) scheduling, for processing queries that are ready. 
(We can alternatively use Earliest Deadline First, for scheduling, where, among the queries whose batches are ready, the query with the earliest deadline is given the highest priority and processed.)


The simulation time progresses as the processing of each query batch gets simulated as shown in Equation \ref{eqn_bet} (Algorithm \ref{algo_generate_schedule}, line 15)
    \begin{equation} \label{eqn_bet}    
        SimuTime = Q_{i}.BET = Q_{i}.BST + Q_{i}.BCT              
    \end{equation}

Once all tuples of a query are processed, the final aggregation of intermediate results is done and the query is removed from \textit{simuQList} as in lines 18 and 19 in Algorithm \ref{algo_generate_schedule}. As each query batch gets scheduled, it is added to \textit{qryBatchSch} (line 17 in Algorithm \ref{algo_generate_schedule}) and the \textit{schStartIndex} gets incremented. 

If all query batches could be scheduled with non-negative slack time until completion, then \textit{GenBatchSchedule} returns true. If the slack time of a query becomes negative, \textit{GenBatchSchedule} returns false. 


Although not shown in Algorithm~\ref{algo_multi_qry_simulate} for brevity, after line 20 of the algorithm,
we also reset the simulation of the earlier query batches prior to \textit{schStartIndex} back to \textit{initNumNodes}, when the nodes for \textit{schStartIndex} become greater than \textit{initNumNodes+1}. This approach minimizes cost by ensuring the additional nodes are provided for those batches only if 
slack time is still found to be negative.



\floatname{algorithm}{Algorithm}
\begin{algorithm}[t]
\caption{\textit{GenBatchSchedule(simuQList[], \textup{\&}qryBatchSch, batchSizeFactor, $C_{max}$, simuStartTime, schStartIndex, numNodes)}} \label{algo_generate_schedule}
\begin{flushleft}
\begin{algorithmic}[1]
\STATE \textbf{Output:} posSlack (true or false) \textit{: denotes whether all query batches completed with positive slack time. } 
\STATE simuTime = simuStartTime
\STATE For each  query in simuQList determine its batchSize based on batchSizeFactor and $C_{max}$
\WHILE{(simuQList.length > 0)}
\FOR{(i=0; i<SimuQList.length; i++)}
\STATE Determine simuQList[i].NextBRT based on batchSize and input rate
\STATE Determine simuQList[i].BCT, simuQList[i].FAT with numNodes with numNodes using the $BatchTime_q()$ and $FinalAggTime_q()$ functions.
\STATE Determine simuQList[i].BST using Equation \ref{eqn_bst} 
\STATE Determine simuQList[i].SQSlackTime using Equation \ref{eqn_slack_single_qry}
\ENDFOR
\STATE \textbf{end for}
\STATE Pick the query, selQry, with the least slack time, from among those which are ready at simuTime. If none of the queries is ready at simuTime, pick the one with the (lexicographically) least value of (NextBRT, slack time).
\IF{(selQry.SQSlacktime < 0)}
\STATE \textbf{return} \textit{false}
\ELSE 
\STATE Update simuTime and selQry.BET  using Equation \ref{eqn_bet} 
\STATE Update pendingTuples in selQry based on batchSize
\STATE Assign attributes of qryBatchSch[schStartIndex] 
 from the corresponding attributes in selQry \{BST, queryID, BST, BET, batchSize, numNodes,
pendingTuples\} \hspace{1em} // Add selQry to  qryBatchSch with attributes \{time, qryID, BST, BET, batchSize, reqNodes, pendingTuples\} 
\IF{selQry.pendingTuples == 0} 
\STATE Update simuTime for final aggregation \& remove selQry from simuQList 
\ENDIF
\STATE schStartIndex++
\ENDIF
\ENDWHILE
\STATE \textbf{end while}
\STATE \textbf{return} \textit{true}
\end{algorithmic}
\end{flushleft}
\end{algorithm}

\subsection{Schedule Optimization} \label{sec_schoptimization}


We now address how to release nodes when not required. 
If the number of nodes required for each query batch in \textit{qryBatchSch} is the same as the initial number of nodes, then the schedule is already optimized as it does not require additional nodes. 

If the number of nodes required for any query batch in \textit{qryBatchSch} is greater than the initial number of nodes, then there can be scope for optimization. For such schedules, we verify their utilization by computing the idle time between the completion and start of consecutive query batches, for batches that were scheduled with a higher number of nodes. 

If idle time exists, we rerun \textit{GenSchedule} from the start of the idle period with the initial number of nodes. If the initial number of nodes is not sufficient, then the appropriate number of nodes required is determined by \textit{GenSchedule} as explained earlier.
This is repeated for each segment with idle time 
before it, and 
the final optimized schedule is obtained by merging the schedules generated for each segment.

In addition, there may be an idle period where no inputs arrive. This period does not overlap with any of the query window durations, i.e. lies outside the window start and end time of all the queries. If such an idle period exists, then the additional nodes, if any, can be released and acquired later when required. In the EMR cluster, 1 primary and 1 core nodes are mandatory, while the task nodes can be released completely. Thus, as part of the schedule determination, we identify the idle period and update the schedule to release all task nodes at the start of the idle duration and acquire them at the end of the idle period. A similar idle period may exist for long running queries where there are sufficient gaps between consecutive query batches scheduled for execution. In a similar manner, task nodes can be released for such scenarios too. 

During simulation, we assume that nodes can be acquired and released instantaneously. However, we address the practical delays encountered in node acquisition and release in Section \ref{sec_schexecution}.

\subsection{Determination of Optimal Batch Size Factor and Initial Configuration} \label{sec_configdetermination}

Function \textit{GenSchedule}() generates the schedule for the given batch size factor and initial configuration. To determine the optimal batch size factor and initial configuration, we run \textit{GenSchedule}() along with optimizations described earlier, for different batch size factors from 1X to some maximum value, say 6X, starting with the different initial configuration, e.g. $Cf_{1}$ to $Cf_{5}$, and choose the configuration and the batch size factor that minimizes the overall cost, while meeting deadlines. 

Note that the minimum batch size can vary for each query, and it is determined as explained in Section \ref{sec_schgeneration}. The simulation can be run in parallel to the query execution to reduce the additional delay due to the simulation.





Running the simulation for all possible batch sizes factors and initial number of nodes may be expensive, and, hence we run the same for a chosen set of batch size factors and initial configurations. 

\fullversion{
The intuition behind the optimality of our scheduling algorithm is explained here. Each simulation run starts with an initial number of nodes and further nodes to are added in steps of one at a time to the minimum number query batches, such that all query batches can be scheduled with positive slack time, thereby keeping the additional number of nodes required and its duration of usage to the minimal. Further, by running the simulation for different combinations of batch sizes and initial number of nodes, the scheduler determines the optimal schedule. 
}

\section{Schedule Execution} \label{sec_schexecution}

In this section, we explain the execution of the schedule generated as part of the simulation. The scheduler carries out node management and scheduling of query batches for processing, based on batch readiness.

The schedule contains the number of nodes required at any given point in time. Although the simulation considers that nodes can be acquired and released instantaneously, cluster platforms require a certain duration to allocate or release a node. Here, we explain how such delays are accommodated. 

The scheduler keeps track of the latest number of nodes requested to the cluster. If the required number of nodes changes at any point in the chosen schedule, the scheduler issues appropriate node resize requests as follows.  

\noindent \textbf{Addition of Nodes:}  If the currently requested number of nodes is less than the required number of nodes, then more nodes have to be added. A new node allocation request may have an associated delay ;  upto 6 minutes delay has been observed on AWS EMR  from request to allocation of a node. To accommodate such delays, the scheduler issues requests for new nodes correspondingly ahead of the time when the nodes are required in the schedule. The time ahead of which the request has to be placed can be tuned based on the cluster requirements. 

\noindent \textbf{Release of Nodes:} Similarly, nodes have to be released if the current number of nodes is greater than the required number of nodes at some point in the schedule.  Actual release of nodes happens when there is no active job running on the nodes.
The release operation takes around 1 or 2 minutes for completion on AWS EMR. Thus, nodes are released provided that these nodes remain idle for a certain duration. As node allocation takes a considerable duration, this idle period is set as at least twice the node allocation time.


Execution of query batches is managed by the scheduler. 
The scheduler checks the currently available tuples for each of the queries and determines the queries whose batches are ready for processing. Then, the scheduler computes the slack time of the queries. The query with the least slack is chosen and scheduled for processing, following the LLF policy. 

On processing the query batch, its result, referred to as an intermediate result, is stored. Once all the batches of the query have been processed, the final aggregation is done on the intermediate results, and the query is removed from the scheduler queue. 

\fullversion{
\floatname{algorithm}{Algorithm}
\begin{algorithm}
\caption{MultiQryScheduleLLF()} \label{algo_multi_qry_scheduler}
\begin{flushleft}
\begin{algorithmic}[1]
\STATE \textbf{Input:} 
\STATE \hskip0.5em $config[]$ $//List of Nodes$
\STATE \hskip0.5em $qList[]$  $//list$ $of$ $queries$
\STATE \hskip0.5em $qListToAdd[]$  $//list$ $of$ $queries$ $to$ $be$ $added$
\STATE \hskip0.5em $qListToDelete[]$  $//list$ $of$ $queries$ $to$ $be$ $removed$
\STATE \hskip0.5em $ipRate$ $//estimated$ $input$ $rate$
\STATE $qReadyListy[]$  $//queries$ $whose$ $batches$ $are$ $ready$
\FOR{$EVER$}
\STATE $qryChanged$ $=$ $false$ $//denotes$ $addition$ $or$ $deletion$ $of$ $query$
\IF{$sListToAdd$ $contains$ $new$ $queries$}
\STATE $Append$ $New$ $Queries$ $from$ $qListToAdd$ $to$ $qList$
\STATE $qryChanged$ $=$ $true$
\ELSIF{$qListToDelete$ $contains$ $new$ $entries$}
\STATE $Remove$ $Queries$ $of$ $qListToDelete$ $from$ $qList$
\STATE $qryChanged$ $=$ $true$
\ENDIF
\STATE $ipRateChanged$ $=$ $false$
\STATE $actIpRate$ = $Compute$ $actual$ $input$ $rate$ 
\IF{actIpRate != ipRate OR qryChanged}
\STATE $schedule$ $=$ $SimulateMain(qList[])$
\STATE $ipRate$ $=$ $actIpRate$
\ENDIF
\STATE $Submit$ $node$ $resize$ $request$ $based$ $on$ $schedule$
\STATE $Clear$ $qReadyList$ 
\STATE $Add$ $queries$ $whose$ $batches$ $are$ $ready$ $to$ $qReadyList$
\STATE $Compute$ $SQSlackTime$ $for$ $queries$ $in$ $qReadyList$ 
\STATE $Sort$ $qReadyList$ $on$ $SQSlacktime$
\STATE $process$ $qReadyList[0]$ 
\IF{$(qReadyist[0]$ $is$ $completed)$}
\STATE $Perform$ $final$ $aggregation$ $for$ $qReadyList[0]$
\STATE $Remove$ $qReadyList[0]$ $from$ $qList$
\ENDIF
\ENDFOR
\STATE \textbf{END}
\end{algorithmic}
\end{flushleft}
\end{algorithm}

The overall logic is given in the Algorithm \ref{algo_multi_qry_scheduler}, \textit{MultiQryScheduleLLF}. Simulation, as shown in lines 7 to 19,  is run whenever a new query is added to the system, or an existing query is removed or when the actual input rate does not match with the estimated input rate. 

Details on handling variations in input rate are explained in Section \ref{sec_variableIpRate}. The function \textit{SimulateMain}, does the simulation and generates the schedule as explained in Section \ref{sec_simulation}. Based on the generated schedule request for acquiring or releasing nodes is issued by the scheduler as in line 20. Lines 21 to 28 in the algorithm, determines the queries which are ready for processing based on its batch size, then computes the slack time. Queries are then sorted based on least laxity and the one with the least laxity is processed. If all tuples of a query have been processed then the final aggregation is carried out and the query is removed from the list of queries, $qList$.}

\section{Handling Variable Input Rate} \label{sec_variableIpRate}

So far, we have assumed that the input data rate can be modeled and the total number of tuples is fixed. 
But there are scenarios where the actual input rate varies from the model. Handling of such scenarios is explained in this section.


In an optimistic model, if the input rate is faster, we assume that tuples arrive early, but the total number of tuples in the window remains the same; if the input rate is slower, we assume that the total number of tuples in the window is smaller. On the other hand, in a pessimistic model, if we see a higher input rate, we assume the higher input rate will continue, resulting in an increase in the number of tuples in the window; if the input rate is slower, we assume that more tuples will arrive later for the same window, and the total number of tuples in the window remains the same. 

Based on the application, the system may choose to use an optimistic or a pessimistic model for future arrivals. For example, in an online shopping scenario, a pessimistic model may be appropriate on a day when the overall sales are more than the estimated input rate and an optimistic model may be better when the sales are less than the estimated one. 




With a pessimistic model, if we see a higher input rate, we rerun the simulation with the increased total number of tuples. If we see slower input rate, we continue with the existing schedule until the input rate increases, which will happen under the pessimistic model, at which point we rerun the simulation.

During runtime, the actual input rate is determined by averaging the total number of tuples received over a given time duration. Then the scheduler verifies if the actual input rate is within the maximum rate supported by the previously determined schedule. If the actual input rate exceeds the maximum supported rate, then the simulation is rerun, and a new schedule is generated. 

The time duration for input rate estimation depends on the time taken for node allocation in the elastic cluster. For AWS, the node allocation time is observed up to 6 minutes; we use a duration of 3 minutes for the estimation of the input rate to allow reasonably fast adaptation to changing rates while being relatively close to the node allocation time.

But if the input rate varies frequently, then the simulation is rerun repeatedly. To avoid this, as part of the simulation, we also determine the maximum input rate that can be supported by the chosen schedule. We rerun the simulation by increasing the input rate by say $x\%$, and the scheduler checks if the previously determined schedule holds good. This is repeated until the maximum rate up to which the chosen schedule is sufficient to process all queries is determined.


Also, for multiple streams, we assume that all the matching tuples from the input streams arrive together. In case tuples arrive late in some streams, late arriving tuples can be handled as follows: in each batch, the late arriving tuples, i.e. tuples whose timestamp is earlier compared to the normal start timestamp of the batch are identified and joined against the previous batches of other streams. 

During simulation for assessing the variations in the input rate that can be supported with the already generated schedule, we consider the same amount of variation in each of the input streams. For example, for assessing the maximum rate, we consider both orders and lineitem to be faster by $x\%$ than the modeled input rate. However, in general, the actual input rate may vary independently. The simulation has to be rerun when at least one of the input rates varies beyond the maximum supported rates.

\section{Partial Aggregation} 
 \label{sec_partial}
So far in the simulation model, intermediate results are aggregated only once at the end of processing the last batch of tuples to get the final aggregated result. Instead, aggregation can be done on intermediate results and then the partial aggregates can be aggregated together after processing the last batch to get the final aggregated results. This can be beneficial for stringent deadline scenarios, since partial aggregation reduces the final aggregation time.

As each batch gets processed, the number of tuples in the intermediate result will be the number of groups in the group by. Hence, partial aggregation can be done over a set of batches, and let the number of batches over which the partial aggregation has to be carried out can be made configurable. For example, when set as 25\%, partial aggregation is carried out after every 1/4th of the total number of batches are completed. 

Let the time taken for doing the partial aggregation be denoted as partial aggregation time, $PAT$, which can be computed using the aggregation cost model of the query. Then the batch end time for batches where partial aggregations are carried out can be computed as in Equation \ref{eqn_bet_with_pat}.
   \begin{equation} \label{eqn_bet_with_pat}  
        Q_{i}.BET=  Q_{i}.BST + Q_{i}.BCT + Q_{i}.PAT    
    \end{equation}

During simulation, in Function \textit{GenBatchSchedule()} we generate a schedule by accommodating $PAT$ in $BET$ computation. The simulation process continues as explained in Algorithm \ref{algo_generate_schedule} and the schedule that incurs the least cost is chosen by the scheduler.

During  query processing, as the number of batches reaches the pre-configured value set for partial aggregation, aggregation of results accumulated so far is carried out in addition to processing the current batch. At the end of processing all tuples, partial aggregates are combined to get the final results.

\fullversion{\subsection{Scheduling for Single Query Case} \label{sec_single_query}

We consider a special static scenario with a single query whose input rate does not vary and briefly explain the concept of scheduling here. 

Initially, we assume that all the tuples can be processed in a single batch and thereby account for no aggregation cost. The last batch can be scheduled not earlier than the window end and the number of tuples that can be processed in the last batch is determined based on the duration between the window end and the deadline. The pending tuples have to be processed prior to the window end, and the batch processing cannot be started until the input available time of the last but one batch is met. Then we schedule the last but one batch, based on the number of tuples which can be processed between its input available time and the window end. Similarly, the pending tuples have to be processed in one or more batches, prior to this. Finally, if the number of actual batches required for processing exceeds the initial assumption, then the schedule is regenerated after incrementing the number of batches in the initial assumption and thereby accounting for the final aggregation cost. We repeat this process for different configurations and choose the one that incurs the least cost. }

\section{Implementation Aspects} \label{sec_implementation}


\begin{figure}
  \centering
  \includegraphics[width=\linewidth]{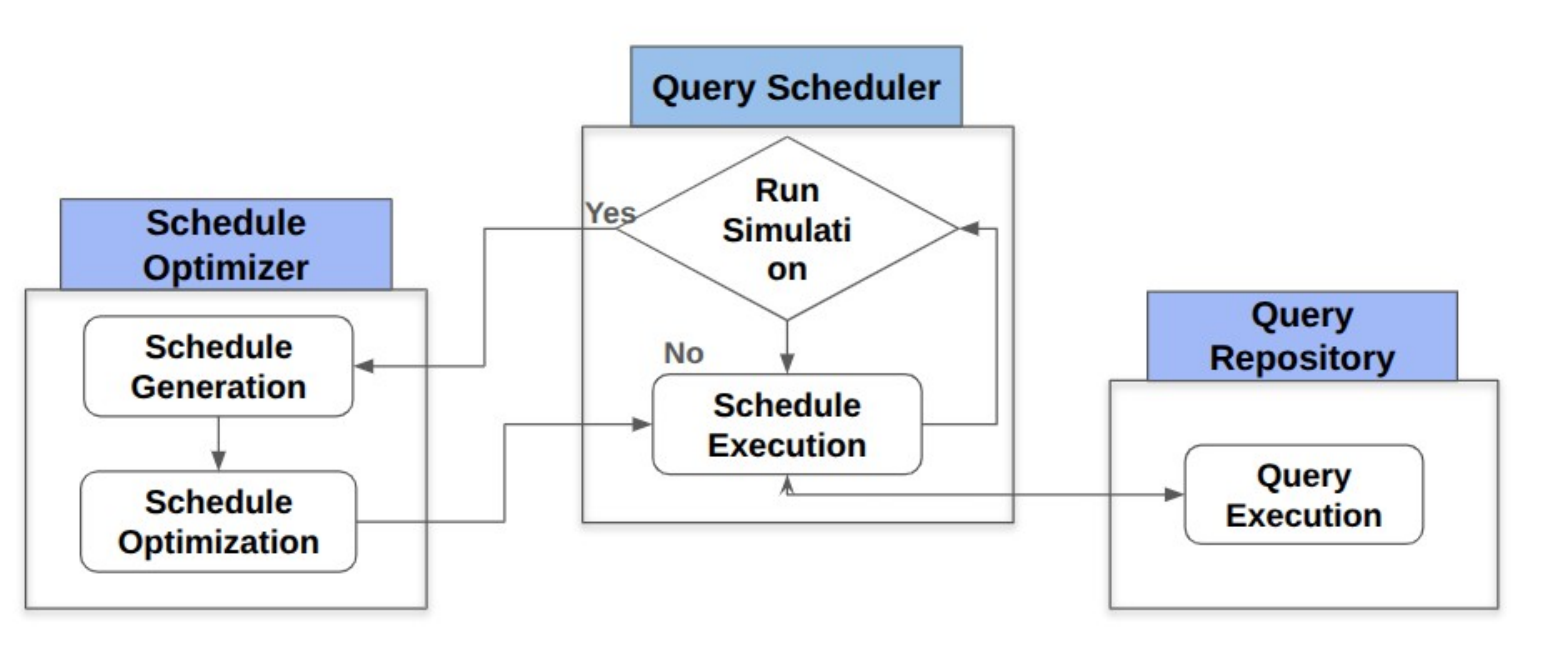}
  \caption{Components of Custom Scheduler}
  \label{fig:customscheduler}
\end{figure}

The scheduling algorithm is implemented as part of a Custom Scheduler module that runs on top of Apache Spark. The Custom Scheduler is made up of three components, namely the Query Scheduler, the Schedule Optimizer and the Query Repository, as shown in Figure \ref{fig:customscheduler}

The primary functionalities of the Query Scheduler are to decide when the simulation has to be run and accordingly invoke the Schedule Optimizer for schedule generation, if needed, and then carry out the execution of the generated schedule.

The Query Scheduler decides to run the simulation by periodically checking for changes in the query requirements, i.e. addition or deletion of queries or input rate variations in the existing queries. 

In addition, the Query Scheduler keeps track of the current number of nodes in the cluster and the actual number of nodes required as per the generated schedule, and issues appropriate node resize requests if needed. 
The Query Scheduler determines the queries whose batches are ready for processing based on the number of tuples received and then computes their slack time or laxity. Then it schedules the query with the least laxity for processing by invoking the Query Repository to process the selected query. 

The Schedule Optimizer does the simulation and generates the schedule, as and when requested by the Query Scheduler. Schedule generation is carried out for different batch size factors and initial node configurations. For each schedule, the Schedule Optimizer optimizes the generated schedule by reducing the number of nodes based on the node utilization. Finally, the optimized schedule with the minimum cost is chosen and passed to the Query Scheduler for execution. 

The Query Repository contains the metadata of the queries to be processed and their Spark operations since the current implementation is based on Apache Spark.  On processing all the tuples, final aggregation is carried out, and the query is removed from the scheduler queue. 

As a new Spark context creation takes around 25 seconds, we create the Spark context once the scheduler is initiated and then reuse the same for all the Spark operations. But we observed that whenever a cluster resize request is made to reduce the number of nodes, the nodes are released only when there is no active Spark context and no active Spark job running. Hence, for our experimental results presented in Section \ref{sec_results_releasenodes}, we modified the custom scheduler to create a new Spark context whenever a query batch has to be processed. This is to ensure that no active Spark context exists during the release of nodes. The additional time taken for Spark context creation was accommodated in the cost model of the queries.

Since batch sizes are chosen to ensure that no batch takes more than $C_{max}$ time for execution, the Query Scheduler performs these tasks periodically within a maximum interval of $C_{max}$ plus the time taken for simulation. 

\section{Related Work} \label{relwork}

Related work can be categorized as non deadline-based and deadline based approaches.

\subsection{Non Deadline-Based Approaches}

Tang et al.\ \cite{ref_iqp} and \cite{ref_thriftyqp} introduce incrementability and intermittent query processing, where a query can be processed in parts, and the partial results can be combined to get the consolidated result. Their approach does not consider deadlines or elastic scheduling, whereas our approach handles deadlines and
determines elastic scheduling for query batches to minimize the cost while 
ensuring deadlines are met. 

Similarly, Shang et al.\ \cite{ref_crocodile_pap}, Tang et al.\ \cite{ref_crocodile_demo}, propose Crocodile DB, which utilizes query slackness to reduce resource consumption. But these works also do not consider deadlines. 

Sotolongo et al.\ \cite{ref_snowflake} describe dynamic tables in Snowflake that generate incremental views periodically based on user input target lags. Their focus is on processing each batch of tuples within a specific 
lag, unlike query processing on a large window with deadlines.  Also, unlike \cite{ref_snowflake}, our approach determines elastic
scheduling of query batches to minimize the cost while ensuring
deadlines are met.

YARN \cite{ref_yarn} is widely used in stream processing engines.  Yarn caters to generic resource management and job scheduling, but does not support deadline based scheduling.

Cheng et al. \cite{ref_adaptiveSchSpark} categorize jobs as independent and dependent ones and schedule them with the aim of improving throughput and reducing latency. Here, there is no  deadline constraint, unlike our problem statement.

Berg et al. \cite{ref_elastic_inelastic_resallocation} talk about optimal resource sharing among elastic and inelastic jobs such that the mean response time is minimised. While we consider a streaming environment where the readiness of query batches must be considered for scheduling, \cite{ref_elastic_inelastic_resallocation} does not consider a streaming environment and hence jobs are scheduled as and when it arrives. Here, all nodes are allotted to elastic jobs when there is no inelastic job in the queue. We also follow a similar scheduling strategy where the current query batch gets scheduled in all the available nodes. However, as part of future work, we plan to have simultaneous execution of different queries on different subsets of nodes in the cluster. While resources are shared among elastic and inelastic jobs in \cite{ref_elastic_inelastic_resallocation}, as part of future work, we plan to share nodes among different elastic jobs i.e. query batches. 

\subsection{Deadline Based approaches}

There are streaming systems that consider deadlines to process individual tuples \cite{ref_tick}. This is different from our use case, where all tuples in the window have a common deadline. 

Given a set of tasks, the general scheduling techniques are EDF and LLF. Here, we use LLF for scheduling the query batches. While our scheduling is based on LLF, 
unlike \cite{ref_hard_real_time_yarn}, we perform batching to minimize the overall cost in addition to meeting deadlines.

Grosbeak \cite{ref_grosbeak} schedules routine analysis jobs in non-peak hours based on the history of resource utilization. The job is processed in batches, which is similar to our approach, but details on scheduling are not discussed in \cite{ref_grosbeak}.

Work has been done in the past to reduce resource consumption in a cluster environment under deadline bound scenarios.  Dimopoulos et al. \cite{ref_justice}, Sidhanta et al.\ \cite{ref_optex} and Islam et al. \cite{ref_dspark} estimate the optimal number of nodes required for meeting the job deadlines based on past execution logs. 

Wang et al.\ \cite{ref_hard_real_time_yarn} describe a scheduling approach in the YARN context, where, in addition to a deadline, a time-varying value density parameter is used for scheduling. Jobs with maximum value density are given the highest priority. 

Compared to \cite{ref_justice}, \cite{ref_optex}, \cite{ref_dspark} and \cite{ref_hard_real_time_yarn}, our work not only ensures meeting deadlines, we also process in appropriate batches, thereby minimizing the node requirement. 

Zhu et al. \cite{ref_multistage_scheduling_cloud} propose scheduling jobs on an elastic platform with the aim of meeting deadlines while minimizing the number of additional virtual machines. Zhu et al. consider jobs to be  made up of several stages or tasks, each with its start time. While this can be compared to query batches in our methodology, where each query batch has its readiness time, batch start and end time, unlike a job stage (in  \cite{ref_multistage_scheduling_cloud}) which has a fixed start time and computation time, the readiness time and the computation time of a query batch varies based on the number of tuples to be processed. 

Our simulation algorithm evaluates for different batch size factors and chooses the optimal batch size factor and initial configuration for query execution. Also, while additional virtual machines are rented out to meet deadlines in \cite{ref_multistage_scheduling_cloud}, we generate a schedule using the simulation model where additional nodes are added to speed up previous query batches, thereby bringing down the node requirement when deadlines are approaching. 


Huang et al. \cite{ref_agamotto}  propose a scheduling algorithm for incremental query processing on cloud platforms with variable pricing policies by minimising cost while meeting deadlines. While we do not consider variable pricing, our approach minimises cost by determining an elastic schedule by adding or removing nodes appropriately, while in \cite{ref_agamotto}, cost reduction is achieved by scheduling jobs when the prices are lower. In addition, while ours is a streaming scenario where inputs arrive continuously, \cite{ref_agamotto} considers the arrival of data at certain points.

Hou et al.\ \cite{ref_dynamic_scheduler} address the dynamic allocation of nodes for the Hadoop platform by monitoring the job completion status at periodic intervals to ensure that the deadlines are met. But they do not consider batching or input rates.

The closest related work is our earlier work \cite{ref_self_ideas}, which 
considers batching and scheduling for intermittent query processing, but on a single node or a fixed set of nodes.
That paper demonstrates the benefits of processing in batches, showing that scheduling queries with EDF and LLF without a minimum batch size may lead to deadline misses.
However, it does not address elastic scheduling, which is a focus of our current work. In addition, we consider the monetary cost in 
a cloud platform and design scheduling techniques to minimise the actual cost incurred.

While there has been prior work in scheduling jobs to meet deadlines, our work of scheduling based on tuple arrival in a streaming system with a simulation model to add or remove nodes based on demand is a unique contribution. To the best of our knowledge, none of the earlier work considers the combination of
batching and elastic scheduling to minimize cost while meeting deadlines. 

\section{Performance Evaluation} \label{sec_results}

We present our performance results 
using queries from the TPC-H benchmark, some custom queries, and the Yahoo streaming benchmark in this section. 



\subsection{Experimental Setup} \label{sec_tpc_dataset}

Our experiments are carried out in the Amazon Elastic Map Reduce (EMR) cluster platform. \fullversion{EMR supports Big Data frameworks such as Apache Spark and Apache Hadoop. EMR can read and write to Amazon Simple Storage Service (Amazon S3) and Amazon DynamoDB. Data can be read from S3 and transformations can be applied in EMR, and the results can be pushed to S3 for stable storage.  

A typical EMR cluster is made up of Amazon Elastic Compute Cloud (Amazon EC2) instances, where each instance is referred to as a node. Each node belongs to one of the following categories
\begin{itemize}
    \item \textbf{Primary node:} This node manages the cluster and coordinates the distribution of data and tasks among nodes for carrying out the processing. Every cluster has a primary node.
    \item \textbf{Core node:} These are nodes with software components that run tasks and store data in HDFS.
    \item \textbf{Task node:} These are nodes with software components that run tasks, but do not store data in HDFS. 
    
\end{itemize}
Amazon EMR price comprises the Amazon EC2 price and the EMR price, which is billed on a per-second basis. 
}We run our Custom Scheduler in the Primary node of EMR. Input data is read from and the results are written to an S3 bucket. 

There are different pricing options available and we chose the On-Demand pricing model, where the user pays on a per-second basis based on the usage. Though the pricing under Spot instances is cheaper, we chose the On-Demand mode since the instance can be terminated by EMR based on the current demand in Spot Instances. \fullversion{Apart from these other options such as Savings Plans are also available where one or three year hourly spend commitment can be made which reduces cost by 72\% compared to On Demand prices.} EMR has provisions to set the cluster size manually or to scale automatically with predefined rules. For our experiments, we scale the cluster based on the generated schedule.

We use a slightly modified version of the TPC-H Dataset, considering a timestamp in each record of Orders and Lineitem, to mimic an input stream.  A total of 25 GB of data in 4500 files is considered, with an input rate of 1 file of Orders and 1 file of Lineitem per second. Each file of Orders and Lineitem is about 1.2 MB and 5 MB, respectively. Other relations, such as Customer, Parts, Parts Supplier, etc are considered to be static information that does not change during query execution.

A subset of TPC-H queries that support incremental processing, including queries with joins, as in \cite{ref_self_ideas}, has been considered. In addition, custom queries, as shown in Table \ref{tbl:custom-queries}, are also considered. 
Similar to \cite{ref_self_ideas}, for experiments, we assume that matching tuples between Orders and Lineitem are within the same input batch. However, stream-
to-stream join with late arriving tuples can be handled as described in Section \ref{sec_variableIpRate}.


\begin{table}[t]
\caption{Custom Queries}
\begin{center}
\raggedright
\begin{tabular}{lp{2.5in}}
\toprule
\textbf{QueryID}&\textbf{Query} \\
\midrule
    \textbf{CQ1} & \textit{SELECT $count(*)$ as totalOrders FROM orders} \\   
    \textbf{CQ2} & \textit{SELECT $count(*)$ as totalItems, partKey FROM lineItems GROUP BY partKey} \\
    \textbf{CQ3} & \textit{SELECT $count(*)$ as totalItems, suppKey FROM lineItems GROUP BY suppKey} \\
    \textbf{CQ4} & \textit{SELECT $count(*)$ as totalOrders, orderPriority FROM orders GROUP BY orderPriority} \\
\bottomrule
\end{tabular}
\label{tbl:custom-queries}
\end{center}
\end{table}



\subsection{Cost Modeling for TPC-H Queries in EMR}  \label{sec_cost_model_exp}

A standard AWS EMR consists of a Primary node and 1 or more Core nodes and 0 or more Task nodes; computation is carried out on core and task nodes, while core nodes can also access HDFS storage.  
We used EMR version emr-6.13.0, and each of the primary, core and task nodes is of type m5.xlarge with 4 virtual cores and 16 GiB memory.  In our performance study, we consider the following configurations. each with a single Primary and Core node, and 1 or more Task nodes: 1P-1C-1T, 1P-1C-3T, 1P-1C-9T, 1P-1C-13T and 1P-1C-19T, where P, C and T represent the number of Primary, Core and Task nodes, respectively. We consider configurations $Cf_1$, $Cf_2$, $Cf_3$, $Cf_4$ and $Cf_5$, with 2, 4, 10, 14 and 20 worker nodes, respectively.  

\fullversion{
\begin{figure}
\centering
\def\svgwidth{0.45\textwidth}
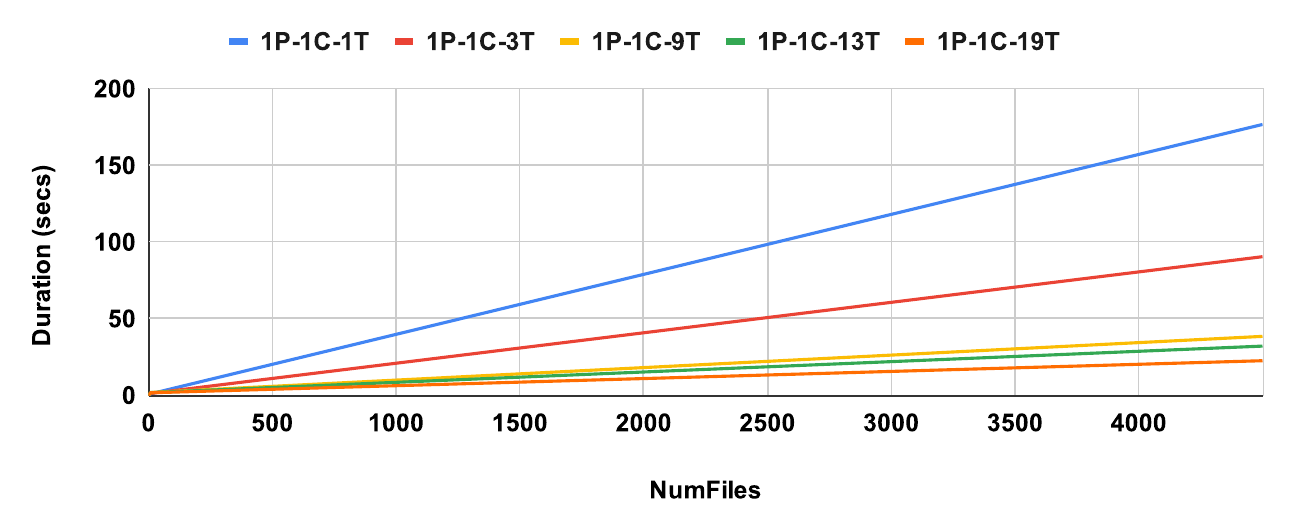
\newline
\caption{TPC-Q1 Processing Duration Vs NumFiles} 
\label{fig:tpc-cost-dur-all-config}
\end{figure}}

\fullversion{
\begin{figure}
\centering
\def\svgwidth{0.45\textwidth}
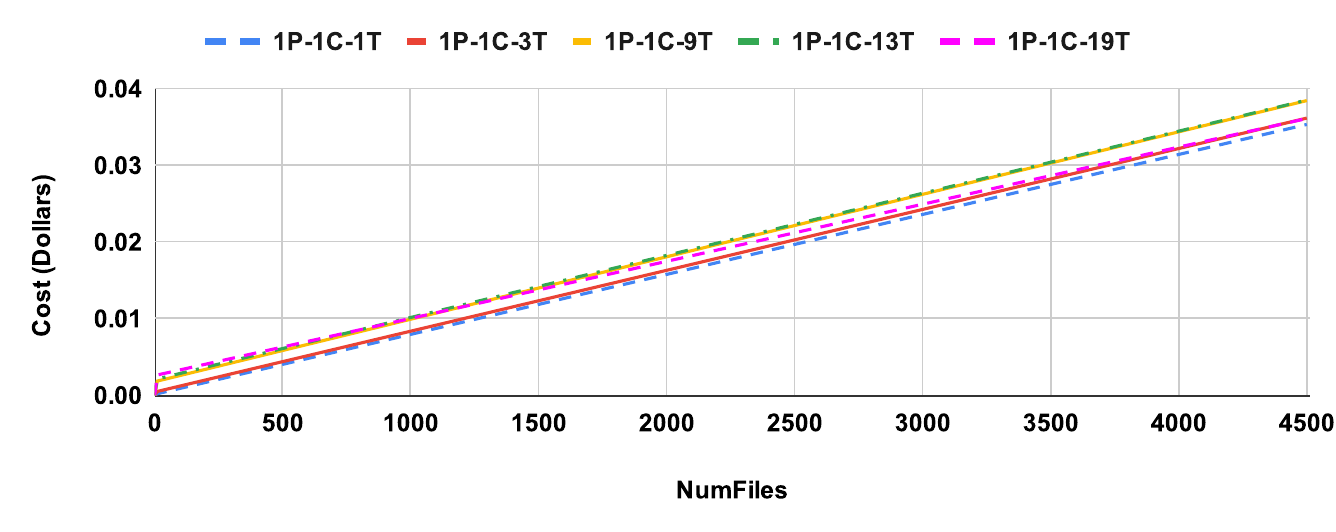
\newline
\caption{TPC-Q1 Cost Vs NumFiles} 
\label{fig:tpc-cost-all-config}
\end{figure}}

For each query, we determine its batch and final aggregation time and cost models for different configurations by measuring time  and monetary cost.


Given that the same queries are executed in a recurring fashion, 
the cost of executing them to fit the cost model gets amortized across a large
number of actual executions.

We experimentally measure the time taken for processing the input by varying in steps of 250 files, where each file is as described earlier, in each run. \fullversion{Figure \ref{fig:tpc-cost-dur-all-config} and \ref{fig:tpc-cost-all-config} show the processing duration required for different input sizes across different node configurations and their associated cost, respectively} Multiple runs were taken, and a cost model, based on the number of files and nodes, was derived by fitting the data using linear regression.
The cost model is linear in the input data size, and also linear on the reciprocal of the number of nodes. Since streaming joins are window based, it is reasonable to assume a linear cost model for queries with joins and the same was observed from our experimental measurements for join queries.

\begin{figure} 
\centering
\def\svgwidth{0.45\textwidth}
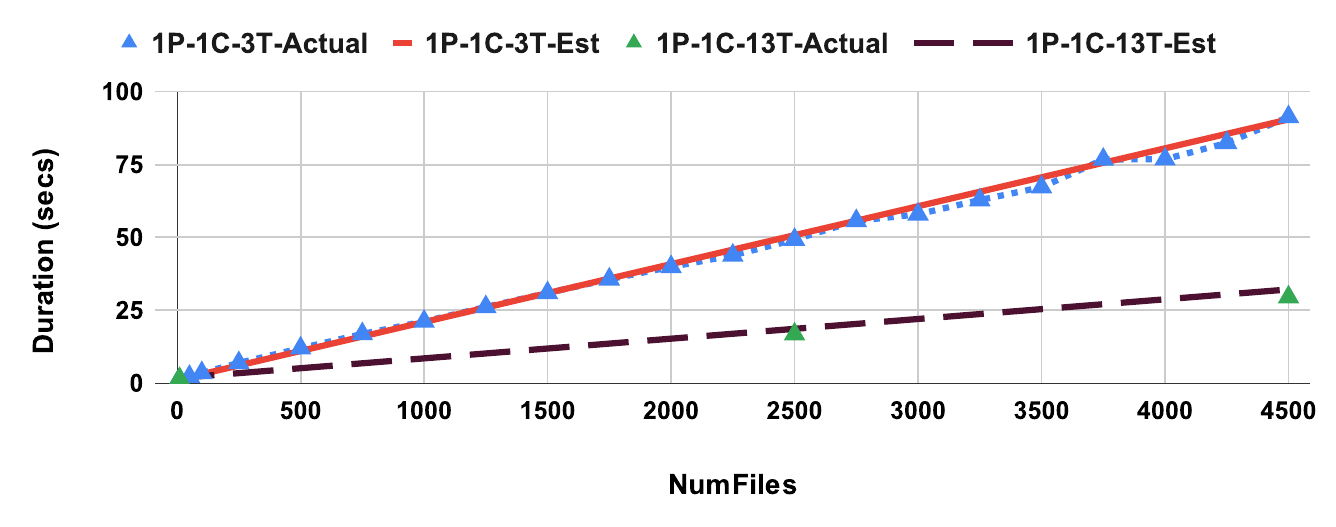
\caption{TPC-Q1 processing duration for different configurations} 
\label{fig:tpc-cost-model}
\end{figure}
Run-to-run variations were observed. For example, processing time measurements for TPC-Q1 had a variance of 4 with respect to the modelled value of 16.5 sec.
Figure \ref{fig:tpc-cost-model} shows the comparison between the actual processing duration and the estimated processing duration for TPC-Q1 using the model by varying the data size, with a fixed C2: 1P-1C-3T and C4: 1P-1C-13T configurations.

To check the accuracy of the model against the configuration, 
we use the actual processing duration for configurations $Cf_1$ and $Cf_3$ to estimate the cost of $Cf_2$, and similarly the cost using $Cf_3$ and $Cf_5$
to estimate the cost using $Cf_4$  

Figure \ref{fig:tpc-cost-model} shows the estimated processing time
for configurations $Cf_2$ and $Cf_4$,
derived as above, against the actual costs using $Cf_2$ and $Cf_4$, across varying input sizes.  Input sizes are measured in terms of the number of files,
each file having around 9500 records.
It can be seen that the time model is quite accurate.


For configurations beyond the maximum number of nodes, i.e., beyond 1P-1C-19T in our case, the processing duration model was obtained by performing a two-step interpolation process, one for the batch size and the other for the number of nodes. First, we fit the processing time duration versus the number of nodes, for a given input size, with a constant plus reciprocal function. 

\begin{figure} 
\centering
\def\svgwidth{0.45\textwidth}
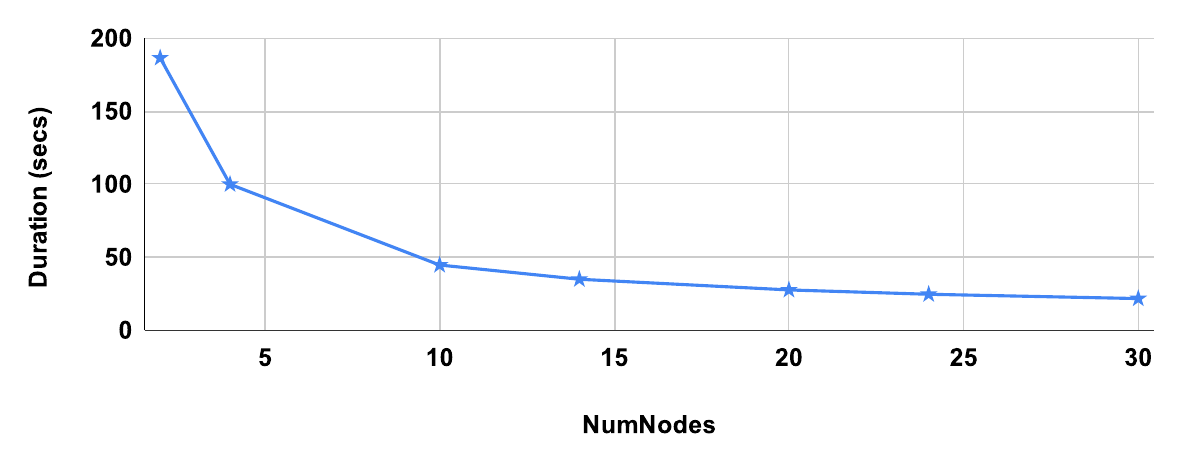
\caption{TPC-Q1 processing duration (for 4500 files) for different number of nodes} 
\label{fig:tpc-cost-model-wrtNodes}
\end{figure}

Figure \ref{fig:tpc-cost-model-wrtNodes} shows the processing duration for 4500 files across different numbers of nodes. Then, using this function, we estimate the processing time for any given number of nodes. With this approach, we have estimated the processing durations for 24 and 30 node configuration.
We also experimentally measured the processing time for selected cases and the actual  measurements were within 25\% difference with respect to the estimated value.


Once the processing duration is determined for a query, the monetary cost incurred can be computed. AWS EMR monetary cost has two components.
The first component is an hourly cost per AWS EC2 node billed on a per-second basis, with a minimum of 1 minute, which depends on the type of compute node.
The second component is a similar per-node cost for using EMR.  For e.g., the hourly rates for Amazon EC2 On-Demand instances and  EMR, for m5.xlarge instances in the Asia Pacific region, are \$0.202 and \$0.048, respectively. Based on the number of instances used and their duration, the total cost is computed. 

\fullversion{
\begin{figure}
\centering
\def\svgwidth{0.45\textwidth}
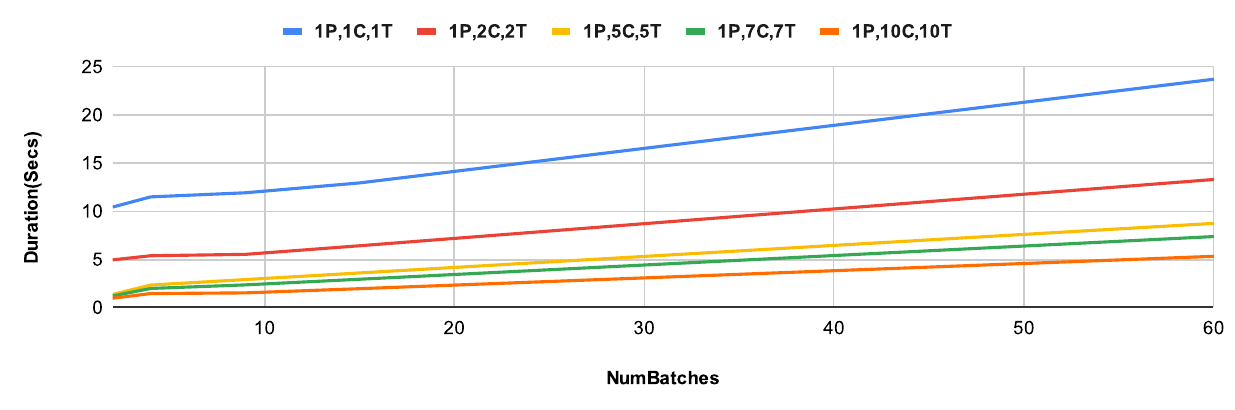
\newline
\caption{TPC-Q3 Aggregation Duration Vs NumBatches} 
\label{fig:q3-agg-duration}
\end{figure}

\begin{figure}
\centering
\def\svgwidth{0.45\textwidth}
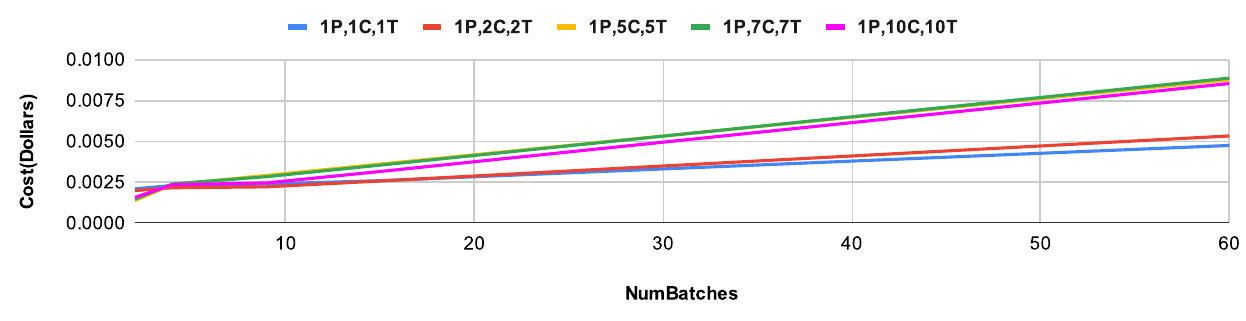
\newline
\caption{TPC-Q3 Aggregation Cost Vs NumBatches} 
\label{fig:q3-agg-cost}
\end{figure}
}

For modeling the aggregation duration, we varied the number of batches to 2, 4, 9, 16, 60, 100, 250 and 450, and measured the aggregation time required under different configurations. Using the measurements, aggregation duration was modeled as a piecewise linear model based on the number of batches and nodes. \fullversion{Figure \ref{fig:q3-agg-duration} shows the aggregation processing duration for different batch sizes. Figure \ref{fig:q3-agg-cost} shows the cost incurred for the aggregation with different batch sizes. It can be observed that As the number of batches increases, the processing duration increases and thereby the cost incurred also increases.} 

\fullversion{In addition, given a set of N queries, it can be laborious to perform measurements for all N queries across all different configurations. For such cases, we can measure all N queries in a selected configuration and compute the relative factor of a query with respect to a baseline query. Subsequently, we can carry out measurements only for the baseline query in other configurations and derive the cost for other queries using the relative factor. For example, using TPC-Q6 as the baseline query, we computed the relative factor for other queries, say TPC-Q1, TPC-Q3 by normalizing the cost for these queries w.r.t. TPC-Q6 cost in 1P-1C-23T and 1P-1C-29T configuration. The relative factors obtained for both configurations are similar.}

\fullversion{
\subsection{Experiments in Single Query Mode}

\begin{figure}[tbh]
\centering
\def\svgwidth{0.45\textwidth}
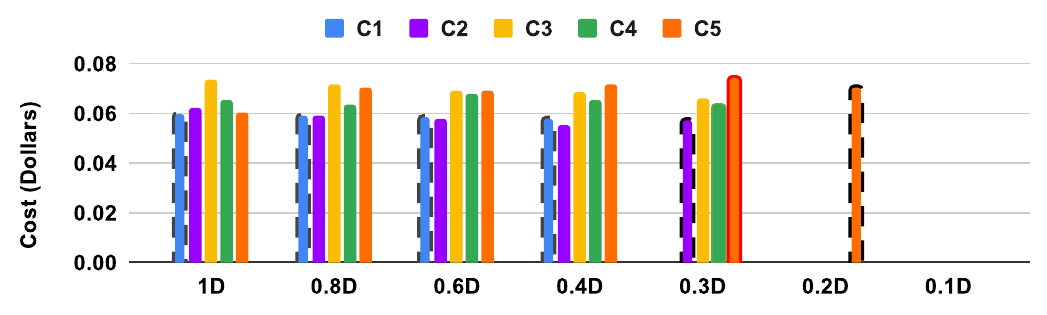
\newline
\caption{TPC-Q3 Cost Comparison in Single Query Scenario} 
\textit{Bars bordered in Black outline denotes the chosen configuration}
\label{fig:single-qry-q3}
\end{figure}

To evaluate the performance of the single query scheduling algorithm, experiments were carried out with TPC-H and custom queries. The time required for processing all the tuples in a single batch starting after the window end time using $Cf_5$ configuration is taken as the query deadline, denoted as 1D. As $Cf_5$ contains the maximum number of nodes, the overall processing duration is the least compared to all other configurations. Thus, for the same deadline, a query which can be executed in a single batch in $Cf_5$, has to be executed in multiple batches in other configurations. Further, the deadlines were reduced to 0.8, 0.6, 0.4, 0.3, 0.2 and 0.1 times the 1D deadline. 

First, the custom scheduler was run in the Compute mode to generate the  schedule for each of the above cases, where the deadline is varied from 1D to 0.1D. Then, the custom scheduler was run in the Execute mode, where the actual query processing was carried out using the pre-generated schedules. From the experiments, the overall time taken for processing is computed. As mentioned in Section \ref{sec_cost_model_exp}, the least granularity in which AWS cost explorer gives the cost summary is in hourly basis.  Since queries were run one after the other in the same configuration, the actual cost from AWS cost explorer will comprise multiple queries. Hence, using the total processing duration from the experiments, the cost incurred was computed using the table \ref{tbl:cost-in-dollars}.    

Figure \ref{fig:single-qry-q3} shows the results with TPC-Q3, where the configuration chosen by the custom query scheduler is given in bounded bars in the figure. The custom query scheduler generated the optimal schedule for all the cases. In addition, schedules were generated for each configuration. TPC-Q3 was executed in each of the configurations. For 0.2D, while a feasible schedule was generated with $Cf_5$, schedule could not be generated for other configurations as query processing could not be completed within the given deadline. For 0.1D, schedule could not be generated in any of the configurations. 

In all cases shown in the figure \ref{fig:single-qry-q3}, queries were successfully completed within their deadline. The configuration chosen by the custom query scheduler is indicated in the figure with bordered bars. Custom query scheduler chooses the configuration which incurs the least cost based on the cost model.

For 0.6D and 0.4D experimental runs show that $Cf_2$ incurs a lesser cost than $Cf_1$. From the experiments, it is observed that the run-to-run variation is in the order of a few seconds with respect to query completion. Hence, though the custom scheduler has chosen $Cf_1$, $Cf_2$ has incurred a lesser cost. The figure \ref{fig:single-qry-q3} also shows that the cost for other configurations $Cf_3$, $Cf_4$ and $Cf_5$ are relatively higher compared to $Cf_1$ except for $Cf_{5}$ in 1D.

\begin{table*}[tbh]
 \begin{center} 
   \caption{Configuration Chosen by Custom Scheduler for Single Query Case}
    \begin{tabular}{ccccccl} 
    \toprule
    \textbf{Query} & \textbf{C1} & \textbf{C2} & \textbf{C3} & \textbf{C4} & \textbf{C5} \\
    \midrule
    \textbf{TCPQ1} & 1D to 0.2D & & & & \\ 
    \textbf{TCPQ3} & 1D to 0.4D & 0.3D & & & 0.2D \\ 
    \textbf{TCPQ4} & 1D to 0.2D &  & & & \\ 
    \textbf{TCPQ6} & 1D to 0.3D &  & & & 0.2D \\ 
    \textbf{TCPQ9} & 1D to 0.3D &  & & 0.2D &  \\ 
    \textbf{TCPQ10} & 0.8D to 0.4D &  & & 0.3D \& 0.2D & 1D  \\
    \textbf{TCPQ12} & 1D to 0.2D &  & & &  \\ 
    \textbf{TCPQ14} & 1D to 0.3D &  & 0.2D & & \\ 
    \textbf{TCPQ19} & 1D to 0.6D & 0.4D \& 0.3D & & 0.2D &  \\
    \textbf{CQ1} & 1D to 0.3D &  & & 0.2D &  \\ 
    \textbf{CQ2} & 1D to 0.3D &  & & 0.2D &  \\ 
    \textbf{CQ3} & 1D to 0.6D &  & & 0.4D \& 0.3D & 0.2D \\ 
    \textbf{CQ4} &  & 0.8D to 0.3D & & & 1D \& 0.2D  \\     
    \bottomrule
    \end{tabular}
    \label{tbl:single-qry-results}
\end{center}
\end{table*}

Similarly, experiments for other TPC-H Queries and Custom Queries were carried out. For larger deadlines, query processing can be scheduled appropriately with a smaller number of nodes. As the deadlines are reduced, more nodes are required to meet the query deadlines. 

The table \ref{tbl:single-qry-results} shows the configuration chosen by our custom scheduler for each query under different deadline conditions. For each deadline as mentioned above i.e from 1D to 0.2D schedule was generated using the custom scheduler. For 0.1D, schedule could not be generated in any of the configurations. 

Then, using the generated schedule with the chosen configuration, query processing was carried out. Runs were taken in other configurations also to verify whether the configuration chosen by the custom scheduler is actually optimal or not. We observed that the custom query scheduler was able to choose the configuration which incurs the least cost in most of the cases. The total number of experimental runs carried out was 78, out of which the custom scheduler chose the optimal configuration for 46 runs. For the rest of the 32 cases, the cost incurred by the custom scheduler configuration was higher, and this was primarily due to run-to-run variations. Among these 32 runs, the percentage difference in cost between the custom scheduler configuration and the one which has incurred the least cost was computed, and it was observed that 24 runs were within 5 \% difference, 3 runs were within 10 \% difference, and the remaining 5 runs were within 18 \% difference. 

Table \ref{tbl:single-qry-results} shows that in most of the cases, $Cf_{1}$ is optimal. Especially for cases with larger deadlines, as the duration available is more, $Cf_{1}$  configuration is sufficient to complete the query processing within deadline as query can be processed in multiple batches starting prior to window end time. Though $Cf_{1}$  configuration requires more time for completing the query processing, in terms of cost it is cheaper than the other configurations. Thus for larger deadlines, query processing can be scheduled appropriately with lesser number of nodes. 

As the deadlines are reduced, more nodes are required for meeting the query deadlines. For example consider TPC-Q19. Here for larger deadline (upto 0.6D), $Cf_{1}$ is optimal. As the deadline is reduced, i.e. for 0.4D and 0.3D, higher configuration, $Cf_{2}$ is required for completing the query processing with deadline. For further stringent deadline of 0.2D, $Cf_{3}$ is required for meeting the deadline. But in terms of the cost, $Cf_{4}$ is cheaper and hence the custom scheduler has chosen $Cf_{4}$. 

Similarly, it can be observed that for 1D of TPC-Q10 and C4, $Cf_{5}$ is the optimal configuration. Here with $Cf_{5}$, both the queries are completed in a single batch, while with other configurations query processing is done in multiple batches and the final results are obtained after aggregation. Though the cost/sec is more for $Cf_{5}$, as the overall processing duration is lesser compared to other configurations, $Cf_{5}$ is cheaper. Thus the custom scheduler picks the configuration which incurs the least cost while meeting the query deadline.}

\begin{table}
 \begin{center} 
   \caption{Simulation Results with Baseline Input Rates}   
    \begin{tabular}{lccccc} 
    \toprule    
    \textbf{Case} & \textbf{INN:2} & \textbf{INN:4} & \textbf{INN:10}  & \textbf{INN:14}  & \textbf{INN:20} \\    
    \midrule
    \textbf{1D:1X} & 1.19:4 & 2.06:4 & 5.13:10 & 6.13:14 & 8.16:20 \\
    \textbf{1D:2X} & \textbf{0.94:2} &  2.05:4 & 5.11:10 & 6.11:14 & 8.13:20 \\
    \textbf{1D:4X} & 0.94:2  & 2.05:4 & 5.10:10 & 6.10:14 &  8.12:20 \\
    \textbf{1D:8X} & 0.94:2 & 2.05:4 & 5.10:10 & 6.10:14 &  8.11:20 \\
    \textbf{1D:16X} & 1.17:4 & 2.07:4 & 5.11:10 & 6.11:14 & 8.12:20 \\ 
    \hline   
    
    \textbf{0.8D:1X} & 1.19:4 &  2.06:4 & 5.13:10 & 6.13:14  &	8.16:20	 \\ 
    \textbf{0.8D:2X} & \textbf{0.94:2} & 2.05:4  &	5.11:10 & 6.11:14 &	8.13:20  \\ 
    \textbf{0.8D:4X} & 0.94:2 &  2.05:4  & 5.10:10  & 6.10:14 &	8.12:20	 \\ 
    \textbf{0.8D:8X} & 0.94:2 &  2.05:4 & 5.10:10 &  6.10:14 &	 8.11:20  	 \\  
    \textbf{0.8D:16X} &  1.52:10 &  2.42:10 & 5.11:10 &  6.11:14 & 8.12:20 	 \\   
    \hline      
    
    \textbf{0.6D:1X} & 1.51:4 & 2.06:4   & 5.13:10 &  6.13:14 & 8.16:20  \\ 
    \textbf{0.6D:2X} & \textbf{1.14:4} &	2.05:4 &  5.11:10 & 6.11:14 &  8.13:20  \\ 
    \textbf{0.6D:4X} &  1.14:4 & 2.05:4 & 5.10:10 & 6.10:14 &  8.12:20  \\
    \textbf{0.6D:8X} & 1.14:4 & 2.05:4 & 5.10:10 & 6.10:14 &  8.11:20  \\ 
    \textbf{0.6D:16X} & 1.52:10 & 2.42:10 &  5.11:10 & 6.11:14 &  8.12:20  \\  
    \hline     
    
    \textbf{0.4D:1X} &  2.77:14 & 2.54:14 & 5.23:14 &  6.13:14 & 8.16:20  \\ 
    \textbf{0.4D:2X} & \textbf{1.55:10} & 2.37:10 & 5.11:10  &  6.11:14 & 8.13:20	 \\ 
    \textbf{0.4D:4X} & 1.56:10 & 2.37:10 & 5.10:10 & 6.10:14 &  8.12:20  \\ 
    \textbf{0.4D:8X} &  1.61:10 & 2.38:10 & 5.10:10 &  6.10:14 &  8.11:20  \\ 
    \textbf{0.4D:16X} & 1.61:14 &  2.52:14 &  5.21:14 &  6.11:14 &  8.12:20  \\  
    \hline
    
    \textbf{0.3D:2X} & 1.91:20 & 2.80:20 & 5.43:20 & 6.32:20 & 8.13:20  \\ 
    \textbf{0.3D:4X} & 1.91:20 & 2.83:20 & 5.41:20 & 6.30:20 &  8.12:20   \\ 
    \textbf{0.3D:8X} & \textbf{1.74:14} & 2.62:14 & 5.20:14 & 6.10:14 &  8.11:20  \\ 
    \textbf{0.3D:16X} & 1.94:24 & 2.85:24 & 5.54:24  & 6.44:24 &  8.23:24  \\  

    \bottomrule
    \end{tabular}
    \label{tbl:cost-comp-config-simu}
\end{center}
\raggedright{
    \textbf{Case}: Deadline:Batch size Factor; Entries: Cost (\$):Max Nodes 
    }
\end{table}

\subsection{Experiments with Baseline Input Rates} \label{sec_results_multiQry_fixedrates}
Experiments were carried out by considering 9 TPC-H queries and the 4 custom queries together. The minimum duration required for processing all tuples in a single batch starting at the window end using $Cf_{5}$ is set as Deadline, 1D. Since all queries are considered together, the deadlines of the queries are staggered such that there is sufficient time to complete the processing of all the queries under 1D scenario. Further cases were generated where the deadlines are reduced to 0.8D, 0.6D, 0.4D and 0.3D.

The results of the simulation are given in Table \ref{tbl:cost-comp-config-simu}. As part of the simulation, the schedule is generated for different batch size factors and configurations. Each case represents the deadline and batch size factor and the columns denote the initial number of nodes (INN). For example, case 1D-1X for INN:2 represents that the deadline is 1D and the initial configuration comprises 2 nodes with Batch size factor 1X. The cost incurred along with the maximum number of nodes required, is specified for each of the cases. For example, 1.19:4 for 1D-1X of INN:2 represents that the cost incurred in simulation is 1.19, while the maximum number of nodes required is 4. For each of the cases, the schedule chosen by the scheduler is denoted in bold in Table \ref{tbl:cost-comp-config-simu}.

From the simulation results, we can observe that as the deadline decreases, the maximum number of nodes required, the batch size factor and the overall cost increase. While a 2 node configuration is sufficient for 1D and 0.8D, additional nodes were acquired for stringent deadline cases. All cases were executed successfully and completed within the query deadlines. 

It can also be noted that the optimal configuration across all cases start with the minimum initial number of nodes, i.e. 2.
Our scheduler detects when more nodes are needed, and thus, there is no benefit to having a greater initial number of nodes.

Table \ref{tbl:cost-comp-config-res} shows the comparison between the simulation and the actual runs.
The initial number of nodes is set to 2, while max number of nodes (MNN) is determined by our scheduler. The batch size factor is chosen based on the minimal cost as per the simulation result in Table~\ref{tbl:cost-comp-config-simu}. As can be seen in Table \ref{tbl:cost-comp-config-res}, the results of the actual runs match well with the simulation results. 

\begin{table}
 \begin{center} 
   \caption{Experimental Results with Baseline Input Rates}
       \addtolength{\tabcolsep}{-2pt}
    \begin{tabular}{lrrrrr} 
    \hline
    \textbf{Case} & \textbf{INN} & \textbf{MNN} & \textbf{BSF}  & \textbf{SimuCost(\$)} & \textbf{ActualCost(\$)}\\
    \hline
    \textbf{1D} & 2 & 2 & 2X &	0.94 & 1.12 \\ \hline
    \textbf{0.8D} & 2 & 2 & 2X  &	0.94 & 1.23 \\ \hline
    \textbf{0.6D} & 2 & 4 & 2X &	1.14 & 1.27  \\ \hline
    \textbf{0.4D} & 2 & 10 & 2X & 1.55 & 1.53  \\ \hline
    \textbf{0.3D} & 2 & 14 & 8X &	1.74 & 1.62\\ \hline
    \end{tabular}
    \label{tbl:cost-comp-config-res}
\end{center}
\end{table}

\begin{table}
 \begin{center} 
     \addtolength{\tabcolsep}{-1pt}
   \caption{Simulation Results with Higher Input Rates}   
    \begin{tabular}{lccccc} 
    \toprule    
    \textbf{Case} & \textbf{2X} & \textbf{4X} & \textbf{8X} & \textbf{16X} & \textbf{32X}\\
    \midrule
    \textbf{2FR:1D} & 4.13:10 & 1.72:4 & 1.42:4	& \textbf{1.34:4}  & 1.93:14 \\ 	
    \hline    
    \textbf{4FR:1D} & - & 7.96:20 & 4.35:10 & 3.44:10  & \textbf{2.99:14} \\ 
    \bottomrule
    \end{tabular}
    \label{tbl:cost-comp-config-higherrates-simu}
\end{center}
\raggedright{
    \textbf{Case}: InputRate:Deadline;  Entries: Cost (\$):Max Nodes  }
\end{table}


\subsection{Experiments with Higher Input Rates} \label{sec_results_multiQry_higherrates}

We ran experiments by increasing the input rate twice and four times, compared to the baseline input rate. The results of the simulation are shown in Table \ref{tbl:cost-comp-config-higherrates-simu}. Here, as we increase the input rate, the total number of tuples to be processed increases. Case $2FR:1D$ column $2X$ denotes the 1D case with 2 times the fixed/baseline input rate with Batch size factor 2X, subsequent columns show cases with larger batch size factors. While the 1D case with the baseline input rate completed with 2 nodes,  the maximum number of nodes required for 2FR and 4FR cases increased to 4 and 14, respectively, as the total number of tuples to be processed has increased compared to the 1D case with the baseline input rate.

\begin{table}
 \begin{center} 
   \caption{Experimental Results with Higher Input Rates}
    \addtolength{\tabcolsep}{-2pt}
    \begin{tabular}{lrrrrr} 
   
    \toprule
    \textbf{Case} & \textbf{INN} & \textbf{MNN} & \textbf{BSF}  & \textbf{SimuCost(\$)} & \textbf{ActualCost(\$)}\\
    \midrule
    \textbf{2FR} & 2 & 4 & 16X &	1.34  &  1.10 \\ 
    \textbf{4FR} & 2 & 14 & 32X  &	2.99 &  2.72 \\
    \bottomrule
    \end{tabular}
    \label{tbl:cost-comp-config-higherrates-res}
\end{center}
\end{table}

All cases were completed successfully within the query deadlines. The results of the actual runs are shown in Table \ref{tbl:cost-comp-config-higherrates-res}.  The simulation results match the actual results reasonably well in terms of  
cost incurred. As the actual runs in AWS are expensive, we have executed some of the configuration and batch size factor combinations as shown in Table \ref{tbl:cost-comp-config-higherrates-res}.

\subsection{Comparison With Other Approaches} \label{sec_results_comp}

We now compare our scheduling algorithm to alternative approaches.

\subsubsection{\textbf{Comparison with Fixed Configuration}} \label{sec_results_fixedConfig}

Here, we compare our elastic scheduling to a fixed configuration, with both approaches using batch size determination as explained in Section \ref{sec_schgeneration}. For each of the cases considered in Table \ref{tbl:cost-comp-fixedconfig-simu}, we carried out a simulation using a fixed configuration.  
In Table \ref{tbl:cost-comp-fixedconfig-simu}, FN represents a fixed number of nodes, while Elastic represents our approach with a variable number of nodes. Table \ref{tbl:cost-comp-fixedconfig-simu} shows the cost incurred with each fixed configuration, provided the deadlines were met successfully. For each of the cases, our approach incurs less cost than the cost incurred by the minimum fixed configuration that meets the deadline. Thus, our approach generates the best elastic configuration to meet the deadline with minimal cost.

\begin{table}[tb]
   \caption{Simulation Results for Cost (\$) with Fixed vs Elastic Configuration}
  \centerline{
    \textbf{Notation:} \textbf{Case}: InputRate:Deadline; 
    \textbf{FN:n} Fixed $n$ Nodes} 
    \centerline{\textbf{Elastic}: Cost : Max Number of Nodes}
 \begin{center} 
  
    \begin{tabular}{ccrrrrr} 
    \toprule
    \textbf{Case} & \textbf{FN:2} & \textbf{FN:4} & \textbf{FN:10}  & \textbf{FN:14}  & \textbf{FN:20} & \textbf{Elastic}\\
    \midrule
    \textbf{1FR:0.6D} & - & 2.05 & 5.11 &  6.11 &  8.13 & 1.14:4 \\ 	
    \textbf{1FR:0.4D} & - & - & 5.10 & 6.10 & 8.11 & 1.55:10 \\ 
    \textbf{1FR:0.3D} & - & - & - & 6.10 & 8.11 & 1.74:14\\
    \textbf{2FR:1D} & - & 2.06  & 5.11 & 6.11 & 8.12 & 1.34:4 \\
    \textbf{4FR:1D} & - & - & 5.13 &  6.13 &  8.15 & 2.99:14 \\
    \bottomrule
    \end{tabular}
    \label{tbl:cost-comp-fixedconfig-simu}
\end{center}
\end{table}

\subsubsection{\textbf{LLF Without Batch Size Determination:}} \label{sec_results_llf}

Our approach uses batching along with LLF based scheduling.
To evaluate the importance of batch size determination, we compare our strategy with using LLF without batch size determination, where the currently available tuples are processed based on the least laxity (minimum batch size is 1 file containing around 9400 tuples, as input is simulated in terms of files/sec). 
We use a fixed configuration for this experiment.

Execution completed within the deadline (1D case) for the 20 nodes configuration and failed for lower configurations.  
The main reason is that the query with the least laxity early
in the execution got executed multiple times with small batch sizes based on available tuples, using up system time,
leaving queries with later deadlines too little time to complete the
execution within the deadline.
In contrast, our approach completed all queries within the deadline with only 2 nodes, as appropriate batch sizes were determined for each query; with our approach, the minimum batch size  was 20 files for TPC-Q12 and the maximum batch size was 218 files for TPC-Q9.

\subsubsection{\textbf{Comparison with EMR Auto Scaling Mode}} \label{sec_results_atuoscale}
We experimented on the FR:1D case with auto scaling enabled, where the cluster size scales automatically using predefined rules. We set the minimum and the maximum number of task nodes as 1 and 29, respectively. Since deadline based rules were not available, meeting deadlines cannot be guaranteed with the auto scaling approach. Further, to compare the difference in terms of cost between our approach and auto scaling, we  defined rules for scaling in and out based on the percentage of YARN Memory available.\fullversion{ as follows: if the percentage of YARN Memory available is less than 15\%, then a node gets added and if the percentage of the YARN Memory available is greater than 75\% then a node gets removed. YARN memory availability was the default option in defining auto scaling rules in the AWS template. Though other options exist based on total load, number of nodes etc. there is no option which based on deadline and hence we defined the scale in and out rules based on the percentage of YARN memory availability.}  

While our scheduler completed the FR:1D case with 2 nodes, auto scaling acquired 10 nodes for the same run. While both approaches completed all the queries within their deadlines, the total cost incurred with AWS auto scaling and our approach is 2.38 and 1.12 dollars, respectively. Thus, our approach is able to meet the deadline at a cheaper cost.  Further, in general, auto scaling gives no guarantees of meeting deadlines.

\subsubsection{\textbf{Comparison with Spark Streaming}}\label{sec_results_streaming}
We experimented using Spark Streaming, considering both TPC and custom queries. Auto scaling was enabled for Spark Streaming to acquire nodes based on the percentage of YARN Memory available. While with our approach, the 1D case completed all queries within their deadlines using 2 nodes, Spark Streaming could not compute the results for queries with joins within the deadline duration. An additional run was taken where only queries without joins were run in Spark Streaming mode. The run completed with 20 nodes, incurring 12 times more cost compared to our approach. Thus, our approach is more efficient in minimizing cost while ensuring deadlines are met.

\subsection{Experiments with Variable Input Rates}  \label{sec_results_multiQry_variablerates}

\begin{figure}
\def\svgwidth{0.48\textwidth}
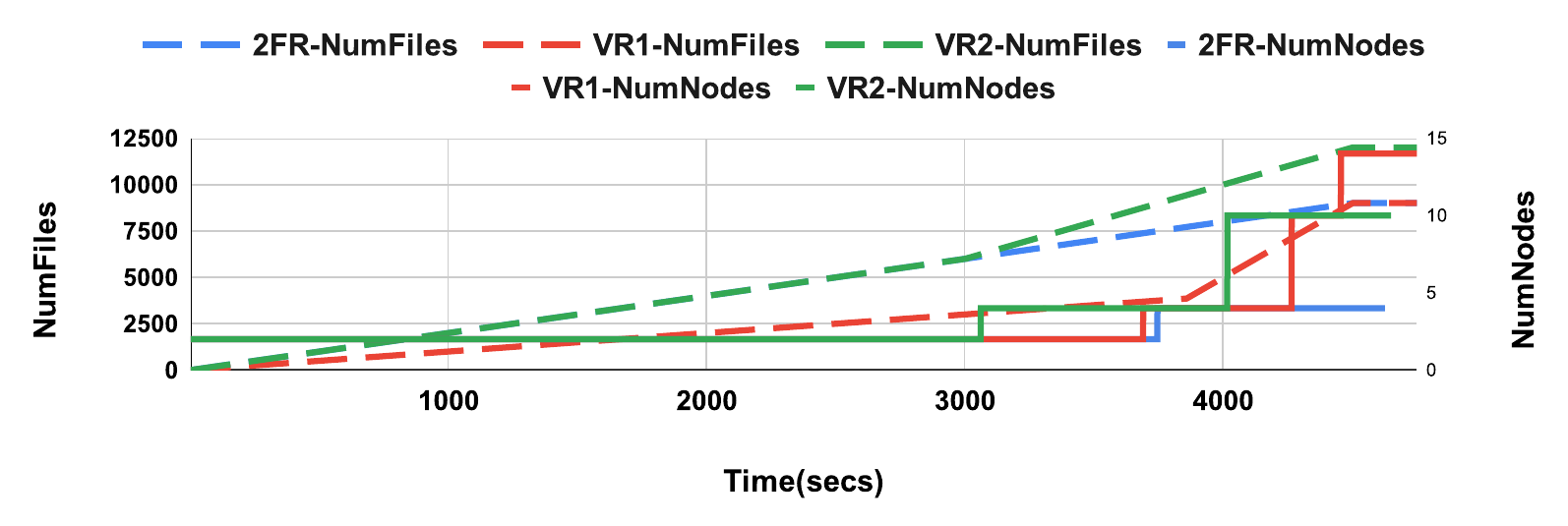
\caption{Variable Input Rate Profiles and Node Requirements} 
\label{fig:multi-qry-varRate-numNodes}
\end{figure}

We also carried out experiments with variable input rates. 
In the first input profile (VR1), the initial input rate was slower than the 2FR input rate and subsequently the input rate  was varied to 8 times the 2FR profile, as shown in Figure \ref{fig:multi-qry-varRate-numNodes}. In the second profile (VR2) the total number of tuples exceeds the maximum number of tuples for the 2FR case. The input rate fed to the simulation was 2FR and the schedule generated was the same as the one shown in Table \ref{tbl:cost-comp-config-higherrates-simu} for the 2FR case. Further, during runtime, the scheduler measures the input rate by averaging the number of tuples received over the last 3-minute duration. If the measured input rate is more than the estimate i.e. 2FR, by 2\% then the simulation is rerun and a new schedule is generated. Further, the new schedule continues until the input rate varies again.

As shown in the Figure \ref{fig:multi-qry-varRate-numNodes}, additional nodes are acquired around 3700 sec for the 2FR case. Though VR1 has a slower input profile compared to 2FR (pessimistic approach assuming tuples to arrive late but by the window end), it acquires additional nodes around 3700 sec, similar to the 2FR case, based on the schedule generated as part of the initial simulation. Further, as the input rate increases around 3800 sec, the simulation is rerun and additional nodes are acquired as per the newly generated schedule. 

For VR2, additional nodes were acquired as the input rate increased, around 3050 sec. Here, as per the initial simulation, which was based on 2FR, additional nodes must be acquired around 3700 sec. But since the input rate has increased, causing an increase in total number of tuples to be processed (pessimistic approach where the number of tuples to be processed increases), the simulation was repeated, and additional nodes were acquired based on the new schedule. 

While 2FR completed with 4 nodes, VR1 and VR2 required 14 and 10 nodes, respectively. Among VR1 and VR2, VR1 requires a larger number of nodes compared to VR2, as tuples arrive late compared to the 2FR profile.



\begin{table}
 \begin{center} 
   \caption{Experimental Results with Variable Input Rates}
    \addtolength{\tabcolsep}{-2pt}
    \begin{tabular}{lccccc} 
    \toprule
    \textbf{Case} & \textbf{INN} & \textbf{MNN} & \textbf{BSF}  & \textbf{SimuCost(\$)} & \textbf{ActualCost(\$)}\\
    \midrule
    \textbf{2FR} & 2 & 4 & 16X &	1.34  & 1.10 \\ 
    \textbf{VR1} & 2 & 14 & 16X &	1.34 &  1.51 \\ 
    \textbf{VR2} & 2 & 10  & 16X &  1.34 &  1.33 \\  
    \bottomrule
    \end{tabular}
    \label{tbl:cost-comp-config-variablerates-res}
\end{center}
\end{table}

Table \ref{tbl:cost-comp-config-variablerates-res} shows the experimental results. 2FR input rates were used for simulation for each of the variable input rate runs. Thus, the initial number of nodes and the batch size factor are the same for all cases 2FR, VR1 and VR2. Further, as the input rates were different, the maximum node requirement and the actual cost incurred varied for variable input rates compared to 2FR. 


\fullversion{As shown in the figure \ref{fig:multi-qry-varRate-numNodes}, as the input rate increases for VR1 and VR3 around 3100 sec, additional nodes i.e. 2 nodes to 4 nodes, have been acquired in both these cases. Further VR3 acquires 10 nodes around 4000 sec as higher rate continues and the total number of tuples to be processed has increased. Hence VR3 incurs the highest cost. VR2 acquires 10 nodes around 3900 sec as its input rate increases only after 3800 sec. VR1 and 2FR acquires 10 nodes around 4100 sec. Further 2FR acquires 14 nodes towards the window end. As VR1, VR2 and VR3 have already acquired higher number of nodes earlier compared to 2FR, additional nodes i.e. 14 nodes were not required towards the window end. Though the maximum number of nodes acquired by 2FR is higher compared to other cases, the cost incurred by 2FR is the least. This is because the pricing for nodes is done based on the duration of usage.} 

Simulation is carried out at the beginning of each run. Further, in each of the variable rate cases, the simulation is rerun when the actual input rate changes from the estimated one, which is after 3000 sec for VR1 and VR3, and 3850 sec for VR2. It was observed that the rerunning of the simulation was triggered multiple times (5 times for VR1 and 9 times for VR2).


\fullversion{To handle the time taken for AWS EMR node allocation and set up, a buffer duration has been configured such that the requests are raised prior to this predefined duration compared to the actual time of requirements. To access the impact of this we took a VR3 run after reducing the buffer duration. Also to access the impact of not rerunning the simulation for input changes, VR3 run was taken in bypass mode where simulation is run only once in the beginning of the run. 

\begin{figure}
\centering
\def\svgwidth{0.45\textwidth}
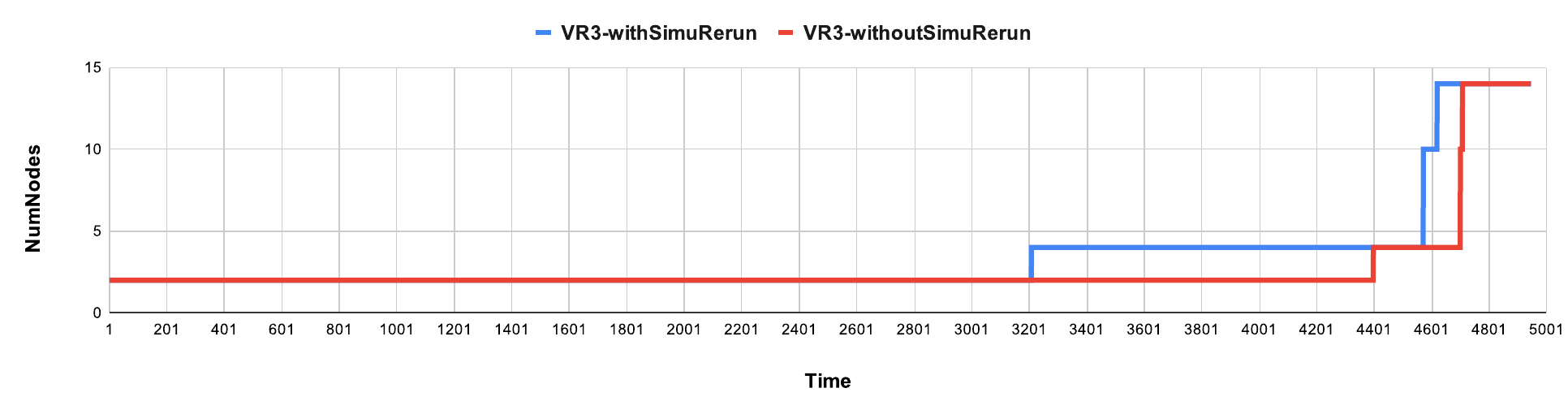
\caption{Comparison with and without rerunning of simulation for Variable input profile} 
\label{fig:multi-qry-varRate-numNodes-withoutsimu}
\end{figure}

Figure \ref{fig:multi-qry-varRate-numNodes-withoutsimu} shows the comparison of how the nodes were acquired for VR3 runs with enabling and disabling of rerunning simulation for variations in input profile. In both the runs the buffer duration for node acquisition has been reduced. Due the reduced buffer duration the total node requirement has increased compared to the result of VR3 run shown figure \ref{fig:multi-qry-varRate-numNodes}. Also all queries failed in the VR3 run without rerunning of simulation. The queries completed at 4946 sec while the final deadline was 4860 sec.}

\subsection{Experiments with Partial Aggregation} \label{sec_results_partAgg}

\begin{table}
 \begin{center} 
   \caption{Experimental Results with Partial Aggregation}
    \begin{tabular}{lrrr} 
    \toprule
    \textbf{Case} & \textbf{Max} & \textbf{Proc. }  & \textbf{Simu}\\
     & \textbf{Nodes} & \textbf{Dur. (s.)}  & \textbf{Cost (\$)}\\
    \midrule
    \textbf{0.4D-NoPartAgg} & 10 & 3115.5276 & 1.55 \\  
    \textbf{0.4D-PartAgg} & 4 & 3690.7754  & 1.15	\\ 
    \textbf{0.3D-NoPartAgg} & 14 & 1693.464 & 1.74   \\
    \textbf{0.3D-PartAgg} & 10 & 2926.7146   &  1.48  \\   
    \bottomrule
    \end{tabular}
    \label{tbl:comp-partAgg}
\end{center}
\end{table}

We carried out experiments with partial aggregation enabled for 0.4D and 0.3D cases with baseline input rate. The frequency of doing partial aggregation was set as 25\% i.e. to carry out partial aggregation once every 1/4th of the total number of batches are processed. 

Table \ref{tbl:comp-partAgg} shows the comparison in terms of node requirement, processing duration and the  cost incurred between runs with and without partial aggregation. With partial aggregation, the maximum number of nodes required has been reduced compared to doing a single final aggregation. This is because the processing requirement after the window end has reduced with partial aggregation. With the reduced node requirement, runs with partial aggregation incurred a lesser cost compared to those with a single final aggregation.


\subsection{Experiments for Release of Nodes} \label{sec_results_releasenodes}

\begin{figure}

\def\svgwidth{0.48\textwidth}
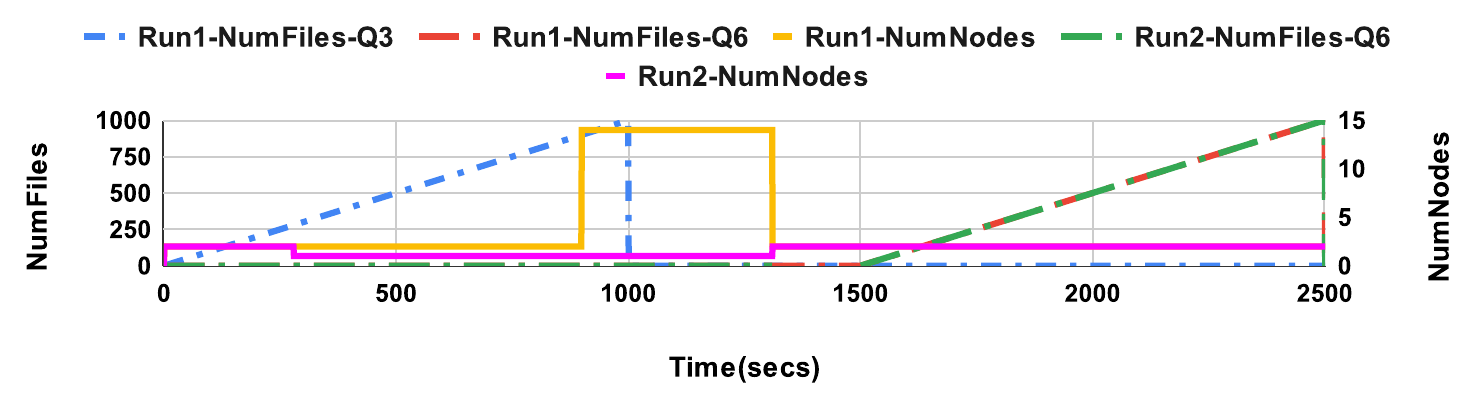
\caption{Release of nodes experiment} 
\label{fig:releasenodes}
\end{figure}

The schedule determined as part of the simulation acquires nodes if needed and subsequently releases them if there is no further requirement for the additionally acquired nodes. To demonstrate the same, two experiments were carried out as shown in Figure \ref{fig:releasenodes}. Run1 has to process 1000 tuples each for Q3 and Q6 with different window start and end points. 

The results show that, as per the generated schedule, additional nodes acquired for Q3 were released at the end of Q3 processing and Q6 continued in 2 node configuration. Run2 has only one query, Q6, whose window start time is 1500 sec. Thus, there is an idle period for the first 1500 sec. Though the scheduler was initiated with the default configuration of 2 nodes, i.e. 1 core and 1 task node, as per the schedule generated, the task node was released during the idle period and later acquired before the window start. 

\subsection{Experiments with Yahoo Streaming Dataset} \label{sec_results_yahoo}


The Yahoo streaming dataset represents an advertisement campaign in which each campaign comprises several events. The mapping between ad-ids and campaign-ids is static data and the same is stored in a CSV file. The query filters the events belonging to the view category and then computes the count of events per campaign, after joining the filtered ad-ids with the static  ad-ids versus campaign-ids mapping info. 

We generated a total of 150 million events with a rate of 40K events per second. In order to avoid the overhead involved in reading data from Kafka, events were generated in JSON files in a designated directory. The scheduler checks the availability of files in the designated directory and processes them based on the chosen batch size.  Similar to the TPC-H dataset, cost modeling and aggregation cost modeling were carried out for different batch sizes in different configurations. 

\begin{table}
 \begin{center} 
   \caption{Experimental Results with Yahoo Streaming Dataset}   
    \addtolength{\tabcolsep}{-2pt}
 
    \begin{tabular}{lrrrrr} 
     \toprule
    \textbf{Case} & \textbf{INN} & \textbf{MNN} & \textbf{BSF}  & \textbf{SimuCost(\$)} & \textbf{ActualCost(\$)}\\
    \midrule
    \textbf{1FR:1D} & 2 & 2 & 32X &	0.75 & 0.83 \\ 
    \textbf{1FR:0.4D} & 2 & 2 & 32X & 0.75 & 0.83 \\ 
    \textbf{1FR:0.2D} & 2 & 4 & 32X & 0.94 &  0.87 \\ 
    \textbf{2FR:1D} & 2 & 2 & 32X  & 0.75 & 0.83 \\ 
    \textbf{4FR:1D} & 2 & 2 & 32X &	0.75 &  0.83\\ 
    \textbf{6FR:1D} & 2 & 10 & 32X & 1.30 & 1.05 \\ 
    \bottomrule    
    \end{tabular}
    \label{tbl:cost-comp-res-yahoo}
\end{center}
 \raggedright{
    \textbf{Case}: InputRate:Deadline}
\end{table}

Experiments were carried out with baseline and higher input rates. The input stream comprised 3750 files with an input rate of 1 file per second, i.e. 40K events per second. The deadline for 1D case was set based on the time taken for processing 3750 files in a single batch using $Cf_{5}$ configuration. Table \ref{tbl:cost-comp-res-yahoo} shows the results. While 1FR refers to the baseline input rate of 1 file per second, 2FR, 4FR and 6FR are the higher input rate cases, which is two, four and six times the baseline input rate, respectively. All cases were completed within the deadlines. The results show that the number of nodes required increases for a strict deadline, 1FR:0.2D and higher input rate case, 6FR:1D, while the other cases were completed with the 2 node configuration.

\section{Techniques to Reduce Simulation Overheads} \label{sec:elasticsch_optimality}

\begin{table}
 \begin{center} 
   \caption{Simulation Results of 2FR-1D Case with Different schStartIndex Settings}   
    \begin{tabular}{lcccccc} 
    \toprule   
    \textbf{BSF} & \textbf{K} & \textbf{INN:2} & \textbf{INN:4} & \textbf{INN:10}  & \textbf{INN:14}  & \textbf{INN:20} \\   
    \midrule
    \multirow{2}{*}\textbf{2X} & 1 & 4.13:10 & 2.87:10 & 5.24:10 & 6.16:14 & 8.17:20 \\  
                                      & 10 & 4.13:10 & 2.88:10 & 5.24:10 & 6.16:14 & 8.17:20 \\ 
                                      & 100 & 4.13:10 & 2.93:10 & 5.24:10 & 6.16:14 & 8.17:20 \\
    \hline
    \multirow{2}{*}\textbf{4X} & 1 & 1.72:4 & 2.06:4 & 5.12:10 & 6.12:14 & 8.14:20 \\  
                                      & 10 & 1.72:4 & 2.06:4 & 5.12:10 & 6.12:14 & 8.14:20 \\
                                      & 100 & 1.78:4 & 2.06:4 & 5.12:10 &  6.12:14 & 8.14:20 \\ 
                                     
    \hline
    \multirow{2}{*}\textbf{8X} & 1 & 1.42:4 & 2.06:4 & 5.11:10 & 6.11:14 & 8.12:20 \\  
                                      & 10 & 1.42:4 & 2.06:4 & 5.11:10 & 6.11:14 & 8.12:20 \\
                                      & 100 & 1.55:4  &  2.06:4 & 5.11:10 & 6.11:14 & 8.12:20\\
    \hline
    \multirow{2}{*}\textbf{16X} & 1 & 1.34:4  & 2.07:4 & 5.11:10 & 6.11:14 & 8.12:20 \\  
                                       & 10 & 1.34:4 & 2.07:4 & 5.11:10 & 6.11:14 & 8.12:20 \\
                                       & 100 & 1.55:4  & 2.07:4 & 5.11:10 & 6.11:14 & 8.12:20 \\
    \hline
    \multirow{2}{*}\textbf{32X} & 1 & 1.78:14 & 2.61:14 & 5.28:14 & 6.16:14 & 8.17:20 \\  
                                       & 10 & 1.93:14 & 2.61:14 & 5.28:14 & 6.16:14 & 8.17:20 \\
                                       & 100 & 1.78:14 & 2.61:14 & 5.28:14 & 6.16:14 & 8.17:20 \\ 
    \bottomrule
    \end{tabular}
    \label{tbl:cost-comp-config-2FR-schIndexvariations-simu}
\end{center}
\raggedright{
    \textbf{BSF:Batch Size Factor; K : StepSize, K in Equation \ref{eqn_schIndexsetting};  Entries: Cost (\$):Max Nodes  }}
\end{table}

\begin{table}
 \begin{center} 
     \addtolength{\tabcolsep}{-1pt}
   \caption{Simulation Results of 4FR-1D with Different schStartIndex Settings}   
    \begin{tabular}{lcccccc} 
    \toprule   
    \textbf{BSF} & \textbf{K} & \textbf{INN:2} & \textbf{INN:4} & \textbf{INN:10}  & \textbf{INN:14}  & \textbf{INN:20} \\   
    \midrule
    \multirow{2}{*}\textbf{4X} & 1 & 7.96:20 & 7.89:20 & 7.87:20 & 7.06:20 & 8.44:20 \\  
                                      & 10 & 7.96:20 & 7.90:20 & 7.88:20 & 7.10:20 & 8.44:20 \\
                                      & 100 & 7.96:20 & 8.12:20 & 7.89:20 & 7.20:20 & 8.44:20 \\
                                     
    \hline
    \multirow{2}{*}\textbf{8X} & 1 & 4.33:10 & 4.07:10 & 5.13:10 & 6.13:14 & 8.15:20 \\  
                                      & 10 & 4.35:10 & 4.07:10 & 5.13:10 & 6.13:14 & 8.15:20 \\ 
                                      & 100 & 4.72:10 & 4.17:10 & 5.13:10 & 6.13:14 & 8.15:20 \\ 
    \hline
    \multirow{2}{*}\textbf{16X} & 1 & 3.42:10 & 3.18:10 & 5.15:10 & 6.14:14 & 8.15:20 \\ 
                                       & 10 & 3.44:10 & 3.20:10 & 5.15:10 & 6.14:14 & 8.15:20 \\ 
                                       & 100 & 3.54:10 & 3.20:10 & 5.15:10 & 6.14:14 & 8.15:20 \\ 
    \hline
    \multirow{2}{*}\textbf{32X}  & 1 & 2.99:14 & 2.96:14 & 5.29:14 & 6.18:14 & 8.19:20 \\
                                       & 10 & 2.99:14  & 2.99:14 & 5.29:14 & 6.18:14 & 8.19:20 \\
                                       & 100 & 2.99:14  & 4.12:14 & 5.29:14 & 6.18:14 & 8.19:20 \\  
    \bottomrule     
    \end{tabular}
    \label{tbl:cost-comp-config-4FR-schIndexvariations-simu}
\end{center}
\raggedright{
    \textbf{Case}: InputRate:Deadline:Batch Size Factor; StepSize: K in Equation \ref{eqn_schIndexsetting}; Entries: Cost (\$):Max Nodes}
\end{table}

\begin{table}
 \begin{center} 
   \caption{Simulation Time Comparison For Different Step Size}
    \begin{tabular}{lrr} 
    \toprule
    \textbf{Case} & \textbf{StepSize:K} & \textbf{SimuTime (sec)}\\
    \midrule
    \multirow{2}{*}\textbf{2FR} & 1 & 1446 \\
                                & 10 & 158 \\
                                & 100 & 70 \\
    \hline
    \multirow{2}{*}\textbf{4FR} & 1 & 2152 \\
                                & 10 & 233 \\
                                & 100 & 126 \\
    \bottomrule                            
    \end{tabular}
    \label{tbl:time-comp-config-res-simu-schIndexvariations}
\end{center}
\raggedright{
   \textbf{BSF:Batch Size Factor; K : StepSize, K in Equation \ref{eqn_schIndexsetting};  Entries: Cost (\$):Max Nodes  }}
\end{table}

The efficiency of our scheduling algorithm in terms of finding the minimum cost and the time taken for simulation is explained here. 

Each simulation run starts with an initial number of nodes and further nodes are added in steps of one at a time to the minimum number of query batches, such that all query batches can be scheduled with positive slack time, thereby keeping the additional number of nodes required and their duration of usage to the minimal. Further, by running the simulation for different combinations of batch size factors and initial number of nodes, the scheduler determines the best  schedule which incurs the minimum cost.

The time taken for the simulation increases when a query's slack time becomes negative and the schedule generated so far contains multiple query batches scheduled one after the other, i.e. without any idle time between the adjacent query batches, thereby increasing the number of times the simulation has to be repeated, until a feasible schedule is generated. Though the simulation can be run in parallel to query processing, simulation time can be reduced by decrementing  \textit{schStartIndex} in steps of $K$, instead of decrementing in steps of 1 (line 15 in Algorithm \ref{algo_multi_qry_simulate}).

The modified step size setting for \textit{schStartIndex} is shown in equation \ref{eqn_schIndexsetting}, where K > 1; \textit{schLength} denotes the length of \textit{qryBatchSch} 

\begin{equation} \label{eqn_schIndexsetting}
schStartIndex=
\begin{cases}
    \textit{if(schStartIndex - schLength) > K } \\
    \; \; \; \textit{schStartIndex -= K}  \\
    \textit{else} \\
    \; \; \; \textit{schStartIndex -= 1 }       
\end{cases}
\end{equation}

Simulation was carried out with values of K as 1, 10 and 100. Tables \ref{tbl:cost-comp-config-2FR-schIndexvariations-simu} and \ref{tbl:cost-comp-config-4FR-schIndexvariations-simu} show the simulation results for different step size, K. We observe that K = 1 gives the minimum cost, but it incurs the maximum simulation time, as shown in Table \ref{tbl:time-comp-config-res-simu-schIndexvariations}.

While K =100 incurs the minimum simulation time, the cost incurred is slightly higher for a few of the cases compared to K = 1 and K = 10. Thus, tuning of K is a trade off between achieving the minimum cost versus reducing the time taken for simulation. 

\section{Conclusion and Future Work}
\label{sec:concl}

Intermittent query processing is ideal for use cases where data over a window period has to be analyzed within some deadlines.
We have proposed scheduling techniques for intermittent query processing in an elastic environment where the processing is done with the appropriate number of nodes to ensure deadlines are met while reducing the overall cost incurred.
Our experimental results establish the benefits of our approach.


Our current scheduling model runs one batch of one query at a time across all available nodes. As part of future work, we plan to extend the scheduling to support the simultaneous execution of different queries on different subsets of nodes in the cluster. Studying the optimality of our scheduling mechanisms is another area of future work. Evaluating our algorithm with dynamic query arrivals is also part of future work.

\clearpage

\bibliographystyle{ACM-Reference-Format}
\bibliography{sample}

\end{document}